\documentclass[11pt]{article}
\usepackage[table,xcdraw]{xcolor}
\usepackage[utf8]{inputenc}
\usepackage[margin=2cm]{geometry}
\usepackage{graphicx}
\usepackage{setspace}
\usepackage{amsmath}
\usepackage{amssymb}
\usepackage{url}
\usepackage{longtable}
\usepackage{pdflscape}
\usepackage{hyperref}
\usepackage{xcolor}
\usepackage{tikz}
\usepackage{xcolor}
\usepackage{tabularx}
\usepackage{booktabs}
\usepackage{makecell}
\usepackage{titlesec}
\usepackage[defaultlines=2,all]{nowidow}
\usepackage{caption}
\usepackage{hyperref}
\usepackage{nameref}
\titlespacing{\paragraph}{0pt}{0pt}{1ex}
\usepackage[margin=2cm,font=footnotesize]{caption}
\usepackage{dcolumn}
\usepackage{rotating}
\usepackage{tikz}

\usepackage[backend=biber,style=authoryear,citestyle=authoryear,isbn=false,url=false,eprint=false,giveninits=true,maxcitenames=1,uniquename=init]{biblatex}

\AtEveryBibitem{
  \clearfield{note}
}

\makeatletter
\newcommand{\setword}[2]{%
  \phantomsection
  #1\def\@currentlabel{\unexpanded{#1}}\label{#2}%
}
\makeatother

\definecolor{bleucite}{RGB}{34,111,212}
\usepackage{hyperref}
\hypersetup{hidelinks}
\AtEveryCite{\color{bleucite}}

\graphicspath{ {./images/} }

\newcommand{\beginsupplement}{%
        \setcounter{table}{0}
        \renewcommand{\thetable}{S\arabic{table}}%
        \setcounter{figure}{0}
        \renewcommand{\thefigure}{S\arabic{figure}}%
     }

\title{From Contact to Threat: A Social Network Perspective on Perceptions of Immigration
}

\author{Yuliia Kazmina$^{1,*}$, Eelke M. Heemskerk$^{1}$, Eszter Bokányi$^{1}$, Frank W. Takes$^{2}$}

\date{{\footnotesize%
    $^1$University of Amsterdam\\
    $^2$Leiden University\\
    $^*$\href{mailto:by.kazmina@uva.nl}{y.kazmina@uva.nl}
}}

\providecommand{\keywords}[1]{\textbf{\textit{Keywords: }} #1}

\begin{document}

\onehalfspacing

\maketitle

\abstract{
Our perceptions are shaped by the social networks we are embedded in. 
Despite the acknowledged influence of close contacts on how we perceive the world, the role of the broader social environment remains opaque. 
Here, we leverage a unique combination of population-scale social network and survey data on perceptions of immigration. 
We find that both direct contacts and a wider social network exposure to migrants matter.
Notably, for natives, network exposure shows a shift from positive to negative association with perceptions of immigration beyond a certain exposure threshold. 
The multi-layer nature of our data highlights this tipping point for next-door neighbors, with private social contexts exhibiting a positive relationship between exposure and immigration perceptions. Furthermore, it shows that contacts spanning multiple contexts also strengthen this relationship.
The provided insights on the interplay between network composition and attitudes toward immigration highlight generic patterns shaping public opinion on pressing societal issues. 
}

\setlength{\parindent}{0em}
\setlength{\parskip}{0.8em}


\section{Introduction}


The challenge of a 'great demographic change' confronts developed countries as drastically declining birth and mortality rates, coupled with perpetual immigration, drive profound demographic shifts \parencite{EC2023}. Persistent immigration and subsequent increasing ethnic diversity in contemporary societies bring forth numerous challenges for members of ethnic minority groups, including discrimination and prejudice \parencite{Nwabuzo2017}. Under these circumstances, prejudice towards ethnic minorities may fuel severe political tensions and societal instability.

We know that people's attitudes towards immigration and prejudice towards ethnic minorities are rooted in patterns of social interaction \parencite{Kunovich2002, McLaren2003}. \textit{Intergroup Contact Theory} (ICT)  argues that prejudice against ‘different others’ can be reduced through close social \emph{contact} between the majority and the minority group \parencite{Allport1954, Pettigrew2006}. An extensive literature has emerged, focusing on how the ethnoracial makeup of one's overall interpersonal environment and, more specifically, one's core networks — groups of individuals with whom one shares emotionally close relationships — impacts attitudes \parencite{Marsden1987, Berg2009, McPherson2006}. Studies consistently demonstrate that the inclusion of migrants within one's social circles, particularly in one's core network, positively influences attitudes towards immigration \parencite{Berg2009, Bienenstock1990, Sanchez2006, Espenshade1996}. 


More recent contributions go beyond the traditional ego-network and dyadic view of ICT in which the emphasis is on the social network surrounding a specific individual or individual connections between pairs of nodes. These efforts aim to expand the scope of meaningful social exposure by considering social network substructures as 'latent social-psychological entities', which are clusters where individuals have a higher probability of being connected and represent significant socialization contexts \parencite{Wölfer2017}. However, results in this body of literature are not straightforward. A recent study 
found that while ICT holds for the close ties, indirect exposure to immigrants through school or place of residence has a (slightly) negative relationship with attitudes \parencite{Bentsen2022}. This lends support to the competing literature of \textit{Group Threat Theory} (GTT) which argues 
that with increasing ethnic diversity in a given area, the predominant ethnoracial group may perceive an erosion of its social, economic, and political influence \parencite{Blumer1958, Quillian1995}. This perceived \emph{threat} prompts members of the dominant group to manifest prejudiced attitudes towards other ethnoracial groups, ultimately leading to intergroup conflict \parencite{McLaren2003, King2007}. 

Navigating between Intergroup Contact and Group Threat Theories poses a persistent challenge in reconciling the relationship between various types of social contact and attitudes toward immigration. 
ICT comes into play in situations of close social proximity, particularly within core networks. These are forms of social contact that ensure not only direct positive contact but also a strong emotional bond. But within a broader social environment that includes a mix of social ties of different types and strength as well as superficial exposure, the strength of the impact of intergroup contact seems to diminish and at one point be overshadowed by the mechanism expected by GTT. So far we have not yet been able to establish the empirical characteristics of this transition point between positive and negative relationship between exposure and perceptions. Addressing this gap is the central challenge 
addressed in this paper. 

We hypothesize that the social environment beyond one's core network provides meaningful exposure to different others that helps facilitate reducing prejudice against immigrants. However, once aggregated at a large scale as recent work predominantly did \parencite{Bentsen2022, Berg2009, Kawalerowicz2021} - an entire school, a municipality, or even a neighborhood, the probability of an observed ego having a meaningful tie or even the slightest exposure to alters is minimal. Therefore we propose to redefine social exposure on a more granular level, in line with established cognitive limits for humans to maintain meaningful social ties \parencite{Dunbar1992}. 

We follow recent empirical work on \textit{Population-Scale Social Network Analysis} and conceptualize a collection of social ties that constitute a representative social fabric one is embedded in as a \textit{social opportunity structure} \parencite{Bokányi2023, Kazmina2023, Soler2024}. This includes a variety of social ties such as people’s household members, family connections, classmates, colleagues, and next-door neighbors. All these potentially meaningful social ties \parencite{Breiger1974} constitute a social opportunity structure, where some ties are activated more often and more intensely than others. Furthermore, these relationships emerge from various social contexts (such as family or work), each offering a distinct set of circumstances, yet collectively forming the complex social environment in which an individual is situated.

The explicit availability of contextual information for each social tie makes it possible to investigate how 
ICT and GTT play out differently in various social contexts. It is to be expected that dynamics of social ties in extended family settings, which were not self-selected, differ from those in, for example, the workplace. We contemplate that exposure to immigrants in these separate social contexts 
would likely manifest varying patterns in an individual's attitudes toward immigration.
For instance, having an immigrant included in the family or living next door presents an opportunity for direct contact of a non-competitive nature, while the work setting might induce a competitive dynamic within the group.

Furthermore, we take into account that previous research showed that multiplexity of social relations, i.e., overlap of roles and affiliations in a social network \parencite{Verbrugge1979}, is very common in core networks \parencite{Small2013, Luken2022}. Nevertheless, it remains unexplored whether having exposure to a minority group in various social contexts 
is associated with reduced prejudice, and if this association is stronger as compared to non-multiplex ties. We hypothesize that experiencing repeated interactions with the same person across different contexts fosters a more nuanced and comprehensive understanding of each other that would be associated with a meaningful reduction of prejudice against minority groups.

While acknowledging the myriad of factors that can influence attitudes towards immigration, including contextual historical and political factors, socio-economic status, and cultural background, among others, our research focuses exclusively on exploring the contribution of social network composition to predicting these attitudes. Therefore, this work aims to compare the importance of various forms of contact in explaining one’s perceptions of immigration. The question that naturally arises is three-fold: \setword{i}{word:rq1}) how do direct close contacts with migrants versus their presence in the wider social opportunity structure explain attitudes towards immigration? \setword{ii}{word:rq2}) how does exposure to migrants play out in different social contexts? \setword{iii}{word:rq3}) how is the multiplexity of exposure to migrants associated with perceptions of immigration? Drawing on the principles of Intergroup Contact Theory and Group Threat Theory, we initially utilize established models to explore how direct close contact is associated with more positive attitudes toward immigration. We then further examine these intergroup relationships within the framework of the wider social opportunity structure. To disentangle the overlapping nature of contextual factors, we separate the different layers of the social opportunity structure. Finally, we account for the multiplexity of exposure to various ethnic groups by examining the contribution of overlapping contact with immigrants in multiple social contexts. 

Our framework for answering the above questions leverages an innovative combination of data sources uniquely available in the Netherlands. 
It provides a comprehensive description of both one’s attitudes as well as one's social networks across multiple social contexts. We source information on perceptions of immigration from the LISS panel, a nationally representative survey covering 5000 households \parencite{LISS}. We also gather information on core self-reported direct close contacts with migrants from this survey. 
We combine this information with social network data encompassing the entire population of the Netherlands, derived from official national registers. \textit{Statistics Netherlands} (Centraal Bureau voor de Statistiek, CBS) synthesized these data to capture a wide array of formal relationships among all residents, including connections through kinship, neighbors, classmates, colleagues, and household members \parencite{Laan2021, Laan2022}. 
Together, these ties constitute an all-encompassing depiction of the social contextual factors, i.e. the social opportunity structure accessible to individuals \parencite{Bokányi2023, Kazmina2023}. Our analysis is based on the 4558 individuals who are both present in the LISS survey and whom we could link to the register-based multilayer network representing the social opportunity structure of the Netherlands in 2018. More descriptives of the data sources are presented in Section~\ref{sec:methods}. 

The Netherlands stands as a pertinent case for understanding how perceptions of immigration relate to social network composition. With its long-standing history of immigration, ranging from guest worker programs in the 1960s to recent refugee influxes, Dutch society exhibits a diverse ethnic composition. More recently, the rise of right-wing parties, often espousing anti-immigrant rhetoric, evident in the 2023 general elections, has further polarized opinions on immigration. More generally, the Netherlands offers insights applicable to Western democracies with a significant share of migrants alongside a dominant group of native subpopulation, as its demographic composition mirrors that of many Western nations. By examining the interplay between perceptions of immigration and social networks in the Dutch context, researchers can draw parallels and generalize findings to other Western democracies facing similar demographic shifts and societal dynamics.

In what follows, we utilize our novel framework to answer the three research questions. We begin by introducing the \textit{Immigration Perception Index} (IPI), which is constructed from a diverse set of questions related to immigration, as sourced from the utilized survey. Next, we quantify \textit{Close contact} and \textit{Exposure} to migrants in the social opportunity structure, from the survey and register data, respectively. We use regression models to show that both have a significant relationship with the proposed Immigration Perception Index. However, for the native subpopulation, the relationship between Exposure and IPI turns concave quadratic, meaning that after a certain threshold, higher exposure is associated with negative changes in attitudes toward immigration within the relevant range of 0 to 1 for Exposure. This pattern is not present for the migrant subpopulation. This suggests that while initial exposure to migrants may foster greater acceptance and more positive attitudes toward immigration among the native subpopulation, there exists a point at which further exposure could lead to negative perceptions, highlighting the importance of understanding the nonlinear nature of the association between social contact and perceptions of immigration.

In addition, we demonstrate that Exposure, particularly within the family and household contexts, i.e. private social contexts, has a positive significant relationship with the Immigration Perception Index, whereas the next-door neighborhood context is the one that reveals a distinct threshold value, indicating a critical point beyond which attitudes toward immigration may shift in a negative direction. Furthermore, when we specifically look at ties in multiple social contexts, referred to as the \textit{Multiplexity} of these social ties, we find that both their presence and quantity have a positive significant relationship with the Immigration Perception Index.


\section{Results}
\label{sec:results}

Below, we provide a concise overview of the essential components comprising the empirical setup, followed by an in-depth exploration of our primary findings. While the research design is extensively elaborated upon in Section~\ref{sec:methods}, the subsequent section briefly introduces how we measure perceptions of immigration as well as exposure to migrants within the social opportunity structure, as detailed in Section~\ref{sec:data_brief}. 
We proceed by juxtaposing the primary outcome variable --- perceptions of immigration --- with the main explanatory variable of interest --- exposure to migrants in Section~\ref{sec:association}. This comparative examination guides subsequent models. Section~\ref{sec:cc_exp} presents regression models aimed at estimating the interplay between both close contact and exposure to migrants with perceptions of immigration, addressing the research question (\ref{word:rq1}). Finally, Section~\ref{sec:contexts} conducts a similar regression analysis, yet segmented by distinct social contexts, to address the research question (\ref{word:rq2}), and emphasizes the role of multiplex ties in addressing the research question (\ref{word:rq3}).

\subsection{Data on perceptions of immigration and social network variables}
\label{sec:data_brief}

To measure the attitudes of survey respondents towards immigration, we create the \textit{Immigration Perception Index} (IPI) based on all immigration-related questions of the LISS survey (LISS Core Study ‘Politics and Values’, wave 11: begin date 03 Dec 2018, end date 26 Mar 2019). 
Higher index values are associated with more positive attitudes towards immigration. See Section~\ref{sec:ipi} for more details on the construction of the index from the various survey items.

In the LISS survey, respondents are also asked to list direct "close contacts with whom they discuss important matters" as well as classify the ethnic background of respective alters.  Based on respondents' answers to this survey question, we create the binary variable \textit{Close contact}, assigning a value of 1 if they report at least one such contact with a migrant and 0 otherwise.

We measure \textit{Exposure} to migrants in the social opportunity structure by calculating the share of people with a migrant background (both first- and second-generation migrants as defined by the administrative registers) among the network neighbors of respondents in the multilayer population-scale network. This network includes family, household, school, work, and neighbor contacts, which together as contexts, or network layers, compose the social opportunity structure of individuals.  

\begin{figure}[ht!]
    \centering
    \includegraphics[width=\textwidth]{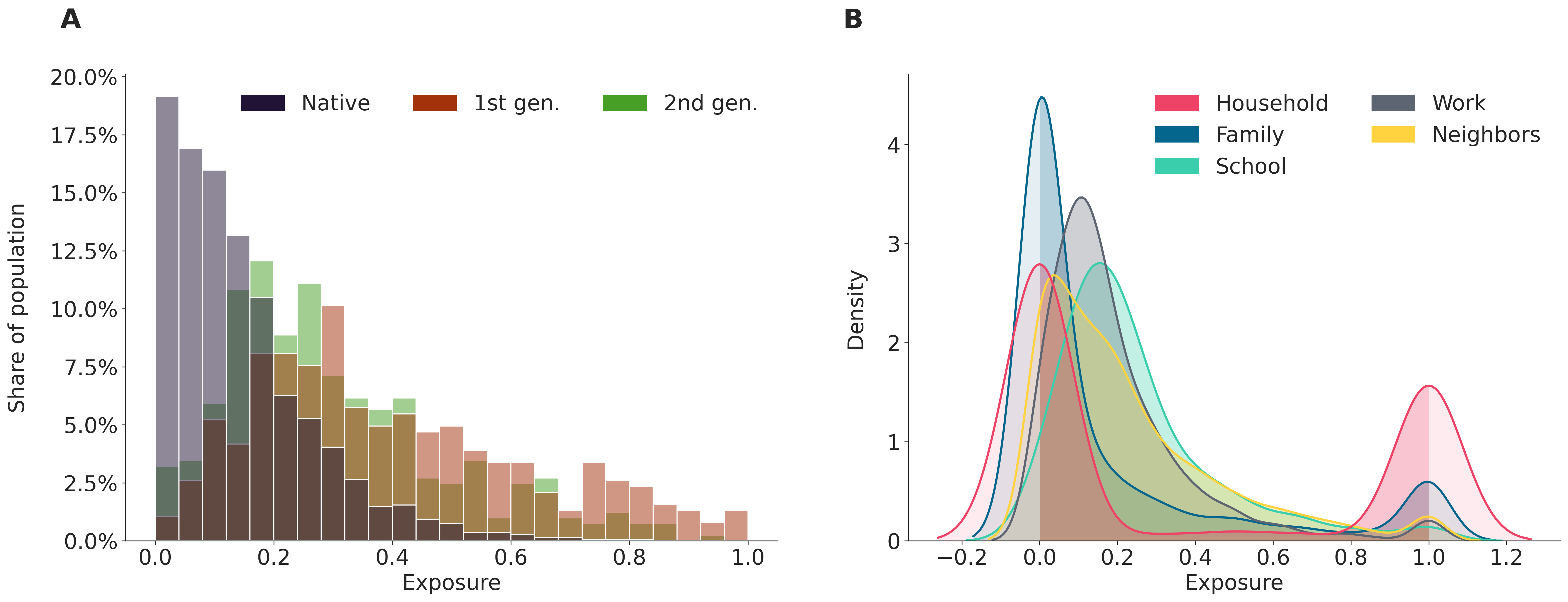}
    \caption{\textbf{Distribution of exposure to migrants in the social opportunity structure of the Netherlands:} (A) Exposure to migrants among natives, first and second generation migrants, (B) 
    Exposure to migrants within social opportunity structure for individual network layers. Values lower than 0 and higher than 1 are artifacts of the kernel density estimation.
}
    \label{fig:fig1_exposure}
\end{figure}

The data reveals notable disparities in the social opportunity structures of natives and migrants. We find that on average, natives have approximately 14\% migrants in their social circles, while migrants themselves exhibit an average exposure of 40\% and 30\% for first- and second-generation subgroups, respectively (Figure~\ref{fig:fig1_exposure}-A). The tails of the distributions indicate a large number of individuals with a migration background who find themselves predominantly surrounded by migrants.

Figure~\ref{fig:fig1_exposure}-B presents a detailed distribution of exposure to migrants for each of the five layers of the network for the whole sample of survey respondents. We find that for all social contexts, the most likely scenario is for a respondent to have either low exposure to migrants ($<$25\%) or extremely high (above 80\%). This pattern is the most pronounced for the family and household layers. For the remaining social contexts such as school, work, and close neighborhoods, extremely high exposure is less common. 

\subsection{Immigration Perception Index and 
Exposure}
\label{sec:association}

\begin{figure}[ht!]
   \centering
   \includegraphics[width=\textwidth]{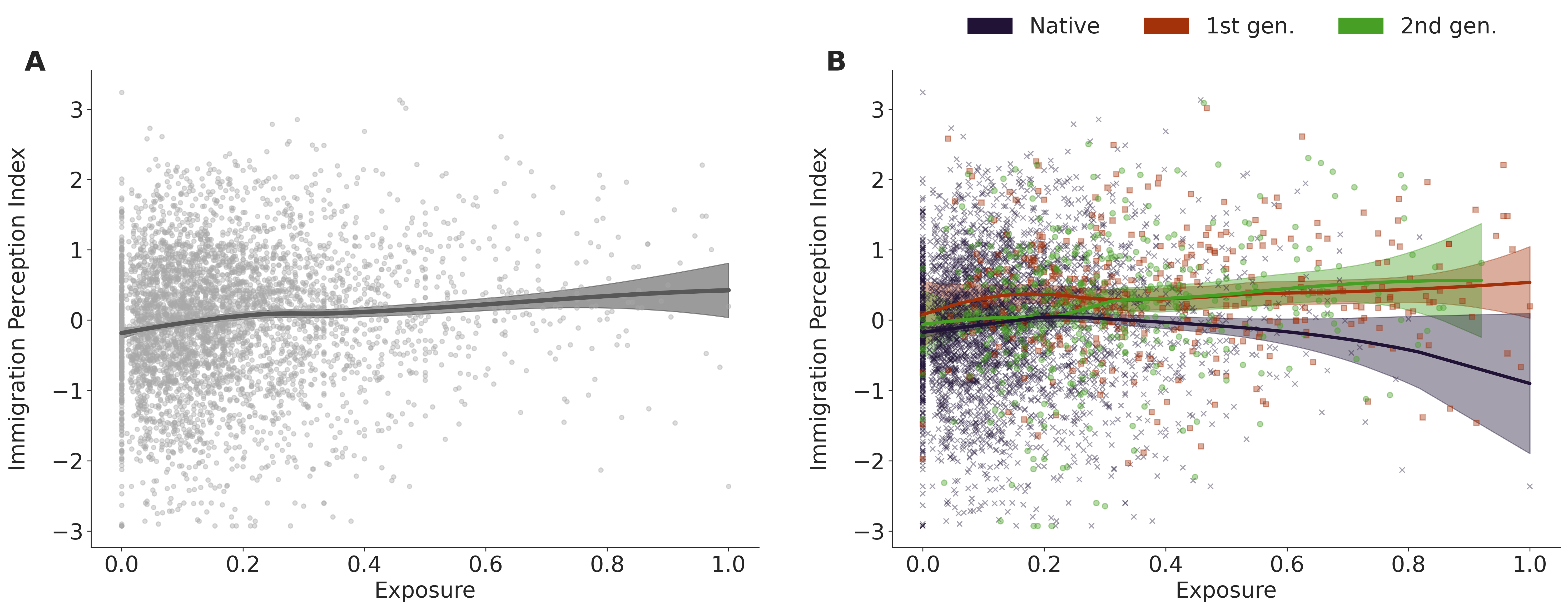}
   \caption{\textbf{Exposure to migrants vs Immigration Perception Index} with the locally smoothed weighting (LOWESS, degree$=2$, fraction$=0.75$) represented by the line plot for (A) the whole sample of respondents and (B) split by migration background. Shaded areas correspond to the confidence intervals.}
   \label{fig:fig2_lowess}
\end{figure}

To explore the relationship between perceptions of immigration and exposure to migrants, we visually depict this association in Figure~\ref{fig:fig2_lowess}-A. The figure illustrates the Immigration Perception Index plotted against the Exposure variable with an estimated nonparametric regression.  While there are plenty of diverse combinations observed, the scatterplot smoothing shows a positive linear relationship between the two with a 95\% confidence interval. The confidence interval remains notably narrow from 0 up to a 50\% Exposure range and then broadens, yet still does not include 0. The observed pattern suggests that as exposure to migrants increases up to roughly 50\%, the relationship with immigration perceptions remains consistently positive. Beyond this point, we observe a more diverse variance of attitudes towards immigration as exposure continues to rise. 

We then uncover that this relationship varies greatly across different subgroups of the population, as defined by their migration background, namely, natives, 1\textsuperscript{st} and 2\textsuperscript{nd} generation migrants (Figure~\ref{fig:fig2_lowess}-B).
For both 1\textsuperscript{st} and 2\textsuperscript{nd} generations of people with a migration background, the results show slightly positive linear relationships that largely overlap. However, for native Dutch people, the relationship between the Immigration Perception Index and Exposure exhibits an inverted U-shaped relationship. At low to moderate levels, exposure to migrants in the social opportunity structure has a positive association with attitudes toward immigration. However, around 30\% this positive association turns negative. This means that in people's social opportunity structure, we see a threshold value, where the positive relationship expected by the Intergroup Contact Theory is outweighed by the negative association expected by the Group Threat Theory. 

\subsection{Both Close contact and Exposure matter for perceptions} 
\label{sec:cc_exp}

To further unpack and quantify the relationship between attitudes towards immigration, close contacts, and exposure, addressing the research question (\ref{word:rq1}), we proceed with multivariate regression analysis. The models are estimated on subgroups of respondents defined by their migration background (natives vs. migrants). As control variables (see Section~\ref{sec:explanatory}) we include age interacted with a binary variable indicating employment status, household income percentile, and highest achieved level of education interacted with a binary variable indicating if a person is in education at the time of data collection. We also include social network degree, i.e., the number of alters in the ego's social opportunity structure. 
Baseline models specifically for the control variables are presented in Supplementary Information~\ref{table:controls}.  

We use four different models to estimate the Immigration Perception Index that use a combination of Close contact and Exposure as independent variables of interest. Model~1 investigates the contribution of the Close contact, and Model~2 explores that of the Exposure variable. Building upon the insights from nonparametric regression estimates indicating a possibility of a non-linear relationship between exposure and perceptions of immigration, Model~3 introduces a quadratic term for Exposure. Finally, Model~4 combines all variables: Close Control, Exposure, and quadratic Exposure for the natives; and Close Control and Exposure for the migrant subgroups. Coefficients for second-generation migrants are estimated with the help of a binary variable indicating second-generation migrants (see Section~\ref{sec:explanatory} for further details).

\begin{figure}[!b] 
    \centering
    \includegraphics[width=\textwidth]{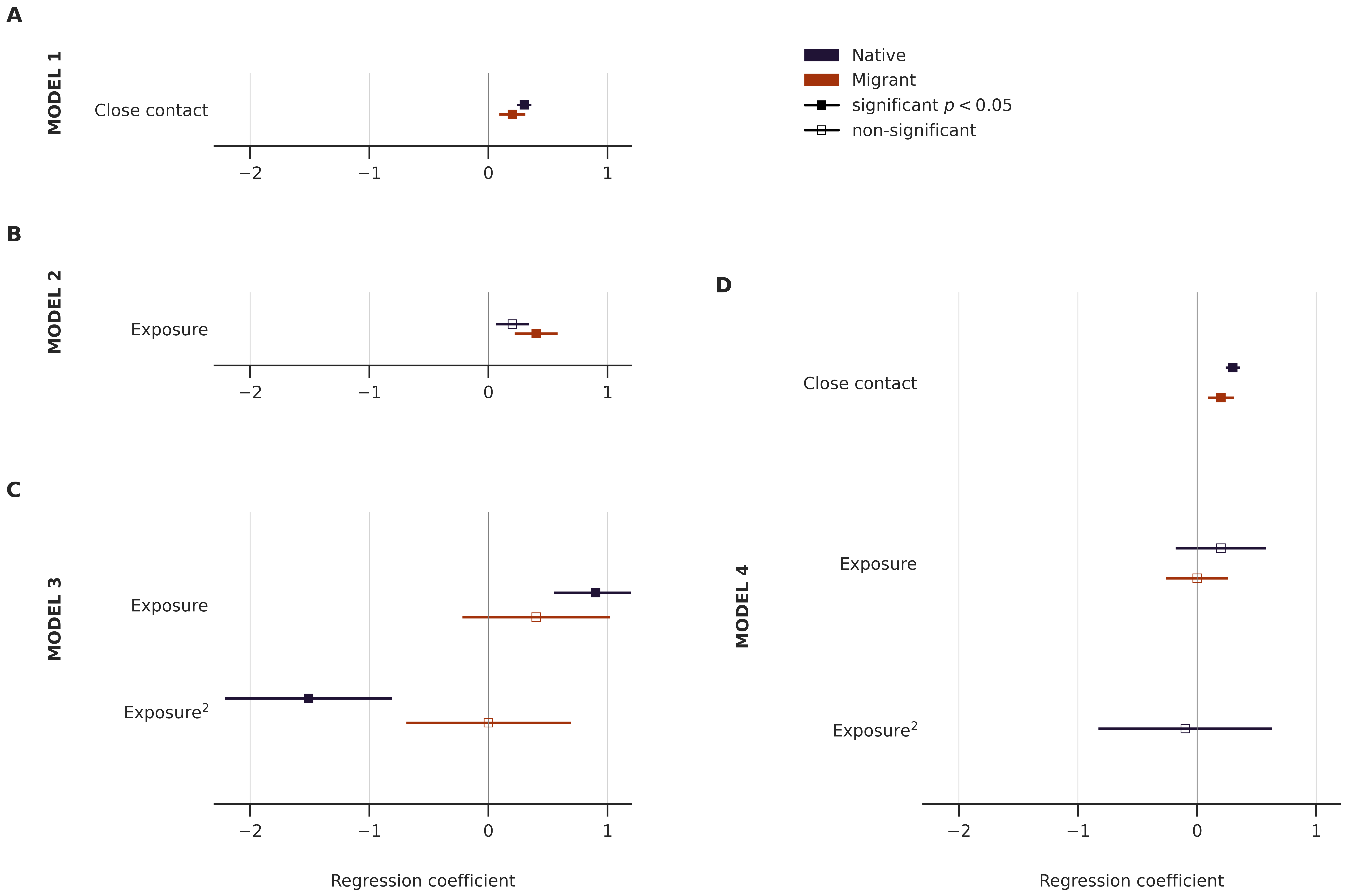}
    \caption{\textbf{Regression results for the Close contact and Exposure variables.} (A-D) Coefficients and their standard errors for Models~1-4 for natives and migrants. Empty markers indicate non-significant coefficients at $p<0.05$, and solid markers indicate significant coefficients.}
    \label{fig:fig3_reg_results}
\end{figure}

Figures~\ref{fig:fig3_reg_results}-A to D depict the estimated coefficients for explanatory variables of interest derived from the regression results for Models~1--4 discussed above. Statistically significant coefficients are represented by solid markers. In Figure~\ref{fig:fig3_reg_results}-A, we find that the coefficient of Close contact is positive and statistically significant for both natives and first-generation migrants, in line with the expectations of Intergroup Contact Theory. This result indicates that people who report at least one direct close contact with a migrant are more likely to have a more positive perception of immigration.

We expect that this positive relationship predicted by ICT is not limited to close contacts alone. Model~2 that uses Exposure as an explanatory variable, at first sight, delivers mixed results. The positive coefficient of Exposure in its linear specification is not significant for the native subpopulation, only for first-generation migrants. As suggested by the nonparametric regression, there is a threshold value for the native subpopulation where the positive association between the perceptions and the exposure turns negative. Therefore, we fit a quadratic regression in Model~3 that confirms the concave relationship for the native subpopulation. Thus, the quadratic Model~3 is a better description of the relationship than the linear Model~2 for the native Dutch subsample. 

Combining both Close contact and Exposure in Model~4 gives a positive significant relationship for close contacts and non-significant for the linear terms of exposure. Thus, close contact is a stronger indicator of more positive views on immigration than exposure. However, when direct close contact is absent (Models~2 and 3), engaging with migrants in a broader social context is also associated with a favorable shift in attitudes toward immigration.

To summarize the significant relationships, and illustrate the threshold values in the nonlinear relationships, we visualize how changes in Close contact and Exposure to migrants are associated with changes in the Immigration Perception index. Figure~\ref{fig:fig5_significant}-A shows the linear relationship from Model~1 (Figure~\ref{fig:fig3_reg_results}-A) with Close contact for natives, first-generation migrants, the quadratic relationship from Model~3 (Figure~\ref{fig:fig3_reg_results}-C) with Exposure for natives, and the linear relationships with Exposure for first- and second-generation migrants from Model~2 (Figure~\ref{fig:fig3_reg_results}-B).

Second-generation migrants do not appear to have a statistically different slope of the relationship between Close contact or Exposure and perceptions of immigration as compared to first-generation migrants. However, the binary variable indicating second-generation status is negative and significant, altering the linear relationship above that of the first-generation migrants, as depicted in Figure~\ref{fig:fig5_significant}-A (see Supplementary Information~\ref{tab:models_sos_migrants} for details). From the quadratic relationship between Exposure and IPI for natives, we can see that there is a turning point after which the increase in the share of migrants in their social opportunity structure is linked to more negative perceptions of immigration. This threshold is at approximately 33\% of exposure.

\subsection{Social contexts and multiplex ties predict attitudes on immigration} 

\label{sec:contexts}

\begin{figure}[ht!]
    \centering
    \includegraphics[width=\textwidth]{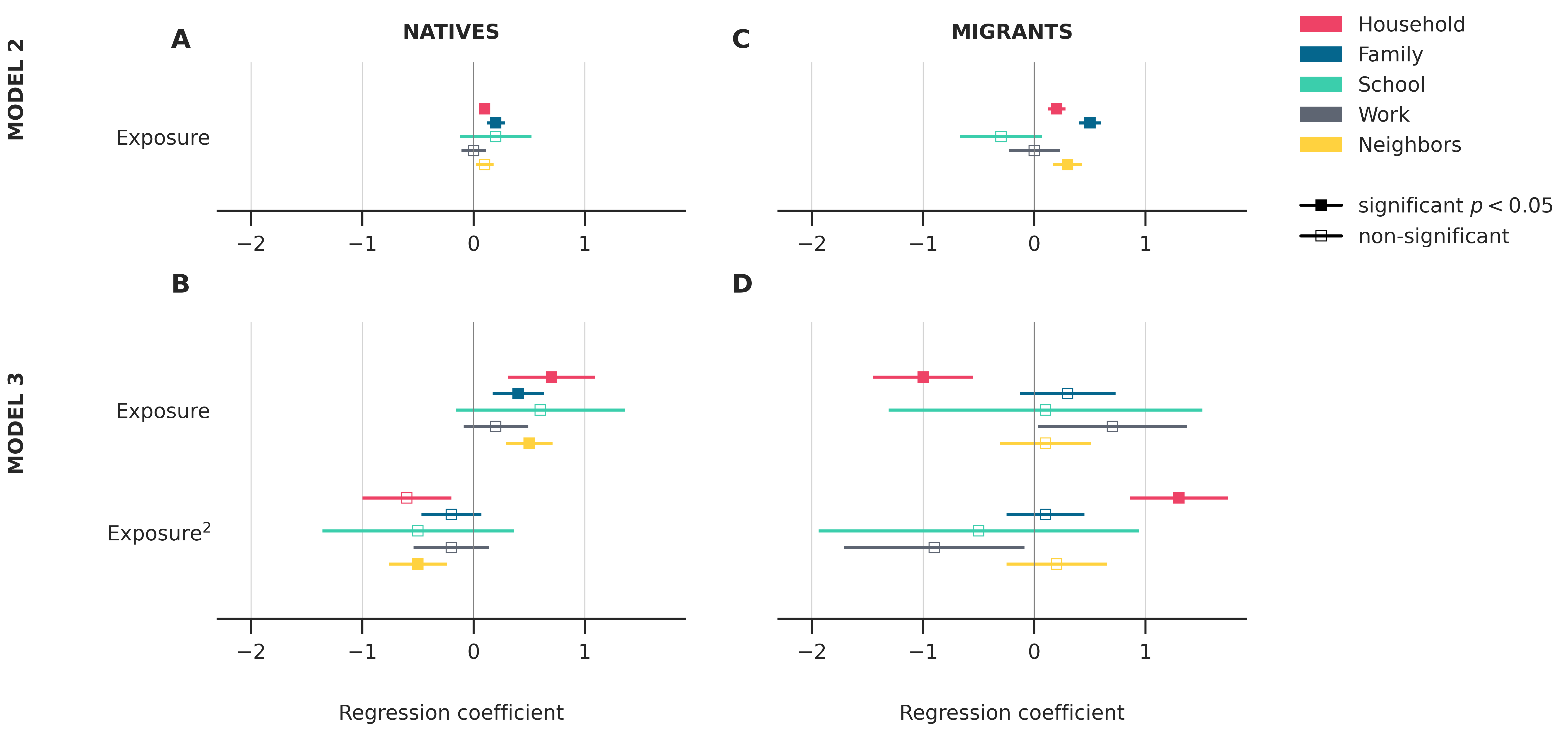}
    \caption{\textbf{Layerwise regression coefficients for Models~2-3 for natives and migrants.}  Empty markers indicate non-significant coefficients at $p<0.05$, and solid markers indicate significant coefficients.}
    \label{fig:fig4_layerwise}
\end{figure}

We now turn to the association between Exposure to migrants in distinct social contexts such as family, neighborhood, household, work, and school and perceptions of immigration as measured by our previously defined index. First, we assess the association between Exposure and the Immigration Perception Index across social contexts, or network layers. Figure~\ref{fig:fig4_layerwise}-A and B illustrate relevant regression results for natives, while Figure~\ref{fig:fig4_layerwise}-C and D showcase the corresponding results for migrants. Both sets of results indicate that exposure to migrants in the household or a wider family, i.e. private social contexts, has a positive significant linear association with their attitudes towards migration for both subgroups of the population. However, this pattern does not extend to exposure to next-door neighbors; notably, for natives, we observe a nonlinear relationship in this context. The quadratic regression demonstrates a concave relationship between Exposure to next-door neighbors and the Immigration Perception Index. This suggests that when natives find themselves in the minority relative to their next-door neighbors, the Group Threat mechanism becomes predominant. For migrants, on the other hand, this relationship remains positive and linear, while also statistically significant. 

\begin{figure}[ht!]
    \centering
    \includegraphics[width=\textwidth]{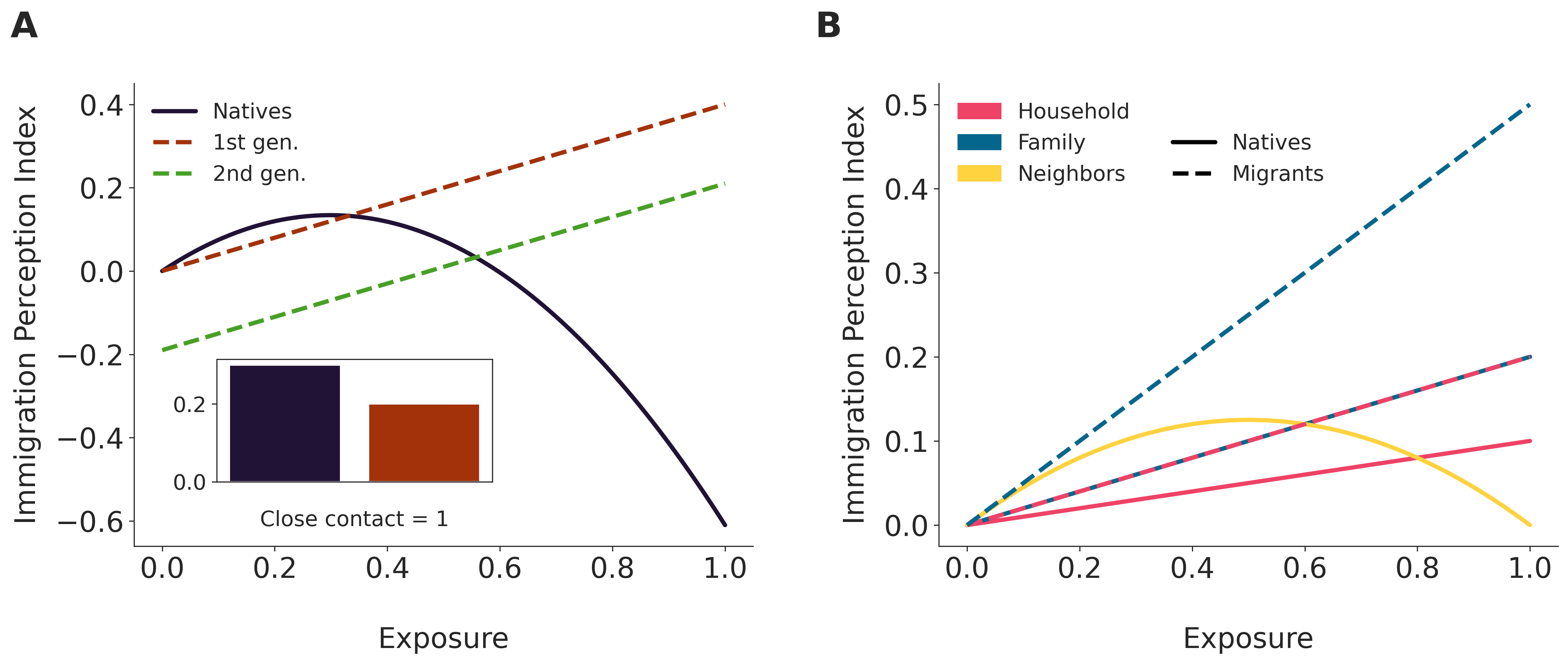}
    \caption{\textbf{Relationships between the Immigration Perception Index and the Close contact and Exposure variables.} (A) Significant relationships for the full social opportunity structure from Models~1--4 ceteris paribus. (B) Per-layer significant relationships from Models~2--3 ceteris paribus.}
    \label{fig:fig5_significant}
\end{figure}

Figure~\ref{fig:fig5_significant}-B depicts how the changes in the Immigration Perception Index are associated with changes in Exposure in the different social contexts based on the significant results of Model~2 and Model~3. For natives, IPI has a positive linear relationship with the share of migrants in the household and family layers. Increasing exposure for natives to migrant neighbors has a positive association until the 50\% threshold after which it turns negative, as shown by the quadratic relationship with the share of neighbors. School and work exposure did not give significant results. For migrants, Figure~\ref{fig:fig5_significant}-B shows a positive linear relationship with family and household exposures, similar to the native subpopulation. Even though the quadratic model gives significant results for the household layer, the distribution of Exposure for the households is bimodal with either very few or most household members being migrants. Hence, within the relevant range of the Exposure variable, there doesn't seem to be a major distinction between linear and quadratic fits; both yield significant results. Therefore, we opt for the linear model based on theoretical assumptions and the nonparametric regression results.

We also hypothesized that multiplex exposure to migrants, i.e., contact with the same alter with a migration background in more than one context, may strengthen the association with perceptions of immigration. For instance, if a migrant is your next-door neighbor as well as a classmate or coworker. We consider two alternative ways of capturing Multiplexity of exposure. First, we include a binary variable indicating if a respondent has a multiplex tie with a migrant in their social opportunity structure. Alternatively, we count the number of such multiplex ties with migrants. We disregard the very common overlap between household and family. The results for both natives and migrants are presented in Table~\ref{tab:multiplexity} (1-2 for natives and 3-4 for migrants, respectively). The multivariate regression analysis shows a statistically significant positive shift if a native has at least one multiplex tie with a migrant. Each additional multiplex exposure to migrants also has a positive association with perceptions of immigration for natives. For migrants, however, none of the considered variations of multiplexity are valid. In fact, multiplex ties with migrants are highly prevalent among migrants themselves, to the extent that this variable loses its predictive power. 

\begin{table}[!htbp] \centering 
\small 
\begin{tabular}{@{\extracolsep{5pt}}lD{.}{.}{-2} D{.}{.}{-2} D{.}{.}{-2} D{.}{.}{-2} } 
\\[-1.8ex]\hline 
\hline \\[-1.8ex] 
 & \multicolumn{4}{c}{\textit{Dependent variable:}} \\ 
\cline{2-5} 
\\[-1.8ex] & \multicolumn{4}{c}{Immigration Perception Index} \\ 
\\[-1.8ex] & \multicolumn{1}{c}{Natives (1)} & \multicolumn{1}{c}{Natives (2)} & \multicolumn{1}{c}{Migrants (3)} & \multicolumn{1}{c}{Migrants (4)}\\ 
\hline \\[-1.8ex] 
 Multiplexity & \cellcolor{bleucite!30}0.34^{*} &  & 0.16 &  \\ 
  & (0.12) &  & (0.16) &  \\ 
  \# multiplex ties &  &  \cellcolor{bleucite!30}0.28^{*} &  & 0.09 \\ 
  &  & (0.08) &  & (0.06) \\ 
  Degree & -0.0002 & -0.0002 & 0.001 & 0.001 \\ 
  & (0.0003) & (0.0003) & (0.001) & (0.001) \\ 

 \hline \\[-1.8ex] 
Observations & \multicolumn{1}{c}{3,750} & \multicolumn{1}{c}{3,750} & \multicolumn{1}{c}{788} & \multicolumn{1}{c}{788} \\ 
R$^{2}$ & \multicolumn{1}{c}{0.07} & \multicolumn{1}{c}{0.07} & \multicolumn{1}{c}{0.05} & \multicolumn{1}{c}{0.05} \\ 
Adjusted R$^{2}$ & \multicolumn{1}{c}{0.06} & \multicolumn{1}{c}{0.06} & \multicolumn{1}{c}{0.03} & \multicolumn{1}{c}{0.03} \\ 
\hline 
\hline \\[-1.8ex] 
\textit{Note:}  & \multicolumn{4}{r}{$^{*}$p$<$0.05; $^{**}$p$<$0.01; $^{***}$p$<$0.001} \\ 
\end{tabular} 
\caption{\textbf{Multivariate regression results estimating the relationship between Immigration Perception Index and Multiplexity} of exposure to migrants in social opportunity structure. Standard errors are clustered at the level of households and presented in parentheses.}
\label{tab:multiplexity}
\end{table} 

\section{Discussion}
\label{sec:discussion}

Persistent and widespread prejudice towards migrants highlights the need for a greater understanding of the underlying factors that play a role in explaining perceptions of migration in a multicultural society. It also raises the need for policies that promote diversity and inclusion, yet take into account their possible adverse effects on perceptions in societies facing major demographic changes \parencite{OECD2020}. To increase our understanding of factors behind the formation of bias against immigration, we investigated how direct contact and wider social exposure to immigrants explain such attitudes. Drawing on the contrasting frameworks of Group Threat Theory and Intergroup Contact Theory, we delved into the specifics of social contexts and the level of exposure at which Intergroup Contact Theory remains applicable, and at what point it gets overshadowed by the adverse dynamics suggested by Group Threat Theory. Moreover, we recognize a possibility of the overlap between various social contexts. Consequently, we conducted a separate examination of multiplex ties --- those that span multiple social contexts --- to determine whether they have a more significant association with altering an individual's stance on immigration compared to simplex ties, which are limited to a single context.

Our findings confirm the principles of Intergroup Contact Theory, demonstrating that direct close contact with migrants is associated with positive attitudes toward them. W
e established that the ICT mechanism also manifests through exposure in the wider social opportunity structure. However, for the native subpopulation, once the exposure in the social opportunity structure reaches a threshold of around 33\%, the association turns negative. This finding suggests that Group Threat Theory should be considered complementary, rather than contradictory, to Intergroup Contact Theory. 

Zooming in on different social contexts, we noted a similar turning point: when natives find themselves in the minority within their immediate spatial neighborhoods, the positive association reverses. Yet, natives who maintain  relationships with immigrants in multiple social contexts are more likely to hold positive views toward migration. These findings on multiplexity align with the importance of close contacts for perceptions since a multiplex tie is more likely to represent a stronger relationship than a simplex exposure in the social opportunity structure. This underscores the nuanced ways in which social context and the nature of ties may influence perceptions of immigration.

While this research has shown the importance of social contextual factors for the formation of attitudes towards immigration, we see it as a step towards a more profound understanding of the relation between the composition of social opportunity structure and perceptions of various matters and immigration specifically. Our study is not without limitations. We test the straightforward relationship between social exposure to migrants and attitudes towards immigration. However, there is a variety of structural constraints that significantly influence social dynamics. First of all, the size and social connectedness of the minority group as compared to the overall network topology, as well as the correlation of node attributes (including views and perceptions) in the population \parencite{karimi2023inadequacy, peel2020, Cinelli2019}. It is to be expected that attitudes towards immigration relate to socio-demographic attributes, as well as a set of other attitudes. For instance, previous work showed how concerns related to the overall state of the national economy and perceptions of the risks influx of migrants may pose to it partially explain individuals' attitudes toward immigration \parencite{Drazanova2022, citrin1997}. Similarly, a consistent positive relationship is observed between the degree of urbanization and more favorable attitudes toward immigration \parencite{Maxwedll2019}. How these different attitudes and characteristics relate to each other and to exposure to migrants in the social world is an important area of future research. We hope our research design shows that such work is now possible. 

By alternating inclusion of explanatory variables of interest while consistently controlling for individual-level characteristics we are able to capture the relative importance of various social network parameters as well as uncover which constituting social contexts are stronger determinants of one’s attitudes toward immigration. The framework and explanatory power of our models can also be used for comparative work, that would give insight into perceptions of immigration outside of the Dutch context. Nevertheless, it is necessary to highlight that with the selected methodological approach we are not yet able to draw causal relationships. The observed relationship between social network structure and perceptions of immigration are associations. To approach causal claims, a longitudinal analysis of perceptions would be required. 

However, when compared to alternative methods of mining social network data, the proposed approach presents several advantages. Previous attempts based exclusively on survey data or online social media data cannot be considered conclusive either as they suffer from certain weaknesses. Surveys provide rather limited information on the social network structure as it is self-reported by an individual and reveals the chosen, very likely most active part of one’s social circle without providing an objective overview of the social environment one is embedded in. On the other hand, online social media leaves plenty of room for mining opinions of interest, for instance, attitudes towards immigration, yet it lacks the capability to distinguish the different types of relationships between individuals \parencite{Drouhot2023}.

Despite its inherent constraints, this study offers valuable perspectives on the reconciliation of Intergroup Contact Theory and Group Threat Theory. It suggests an optimistic view that the benefits of fostering positive attitudes towards migrants extend beyond direct contact, encompassing broader social exposure. It illuminates a nuanced pathway whereby initial exposure to immigrant groups can significantly mitigate negative sentiments toward them. However, our results also delineate a critical threshold beyond which the dynamics of Group Threat become predominant, overshadowing the positive association predicted by Intergroup Contact. Moreover, we show how these thresholds play out differently across social contexts. This observation highlights the complex interplay between increased exposure to minority groups and the activation of perceived threats, suggesting a delicate balance that policy and community initiatives must navigate to foster harmonious intergroup relations effectively. Such policies could include school admission design, community-building programs, intercultural education programs, and initiatives that encourage at least minimal exposure to minority groups and the formation of multiplex ties across different social contexts. By fostering environments where at least minimal exposure to immigrants is guaranteed and multiplex relationships can thrive, we can enhance the positive relationship between diverse social ties and perceptions of immigration. At the same time, policies aimed at integrating migrants should be sensitive to the pre-existing social fabric of communities, ensuring that exposure to diversity does not inadvertently reinforce negative perceptions.


\section{Methods}
\label{sec:methods}

This section discusses the characteristics of the employed survey and network data and the linking procedure (see Section~\ref{sec:data_sources}). Furthermore, it elaborates on the construction of the main dependent variable - the Immigration Perception Index (IPI) in Section~\ref{sec:ipi}, and outlines the explanatory variables of interest along with control variables in Section~\ref{sec:explanatory}. The section is concluded by the description of the steps of the statistical analysis in Section~\ref{sec:stats}. 

\subsection{Data sources} 
\label{sec:data_sources}
The study is based on two main data sources from the Netherlands: the survey for the attitudes and close contact network and administrative register data used to construct the social opportunity structure; below we discuss both, as well as how these two datasets have been linked. 

\textbf{Survey data.} The survey utilized in the study is a nationally representative random-probability survey, namely, the LISS (Longitudinal Internet Studies for the Social Sciences) panel administered by Centerdata (Tilburg University, The Netherlands) \parencite{LISS}. The survey is conducted on a representative sample of 7500 individuals (the minimum age of respondents is 14) and covers various perceptions topics such as gender roles, migration and ethnic minorities, political affiliation, and voting behavior. It also covers respondents’ self-reported close contacts and their background. More specifically, for the purposes of this study, we leverage one of the core studies of the LISS panel, namely, Politics and Values, wave 11, dated April 2019. In this wave of the core study, there are 6439 selected household members. The response rate in the wave is 87.6\%, and completeness is 79.5\%. 

\textbf{Social network data.} The social network data as well as corresponding node attributes are sourced from the Dutch population registers and were provided by Statistics Netherlands (Centraal Bureau voor de Statistiek, CBS). From these various registers, CBS has derived individual-level information on the 1.3~billion formal social ties of all 17.2~million residents as of October 2018~\parencite{Laan2021, Laan2022}. This creates a multilayer population-scale network of social opportunities. In this network, nodes are the people, and connections between them are household, family, neighborhood, school, and work ties, all considered as distinct network layers. Complementary node attributes include equivalised combined household income, as well as individual-level age, highest achieved level of education, migration background, and migrant generation (for non-natives).

In the data, two people are considered household members in case they live together and act as a joined economic unit \parencite{Witvliet2012}. These are in the overwhelming majority of cases either single-person households, partnerships, or family members living together registered at the same address. Kinship ties are derived based on parent-child information and include the following link types: parents, children, siblings, grandparents/grandchildren as well as aunts/uncles and cousins. School ties are derived for people who are admitted to an educational institution at the time of data collection and are distinguished for different levels of education (primary, secondary/specialized secondary, vocational, and higher education). Two people are considered to be classmates if they attend the same institution, in the same year and location, with an equal duration of studies and identical type of educational track. In the work layer, all employees of an organization are connected. For larger organizations (above 100 employees), each employee is connected to the 100 geographically closest colleagues as defined by their place of residence. Next-door neighbors are defined as connections to all members of 10 geographically closest households.
The network is unweighted. If two people work together, and they are also neighbors, the respective pair of nodes share two links: one work link and one neighbor link; we refer to such a link as a multiplex tie. 





\textbf{Data linkage.} Both the survey and the network dataset use a shared unique identifier for each person that allows for the linking of the two. Both data sources are pseudonymized before conducting the research and do not contain personal identifiable information (such as name, precise address of residence, or date of birth). Due to the sensitive nature of the information, all storage, linking, and analyses have been carried out within the secure server environments of Statistics Netherlands. Further information regarding data generation can be found in \cite{Laan2021, Laan2022}. In the combined dataset used in the study, there are 4558 respondents for whom we were able to link survey data on perceptions of immigration to their population-scale social network.

It is essential to underscore that integrating a unique blend of two distinct data sources introduces challenges. While survey data presents self-reported information on close ties, register data provides an objective snapshot of the current social environment of an individual. Self-reported social network information is time-independent, as a listed close contact may have originated from a social context that has since become obsolete. For instance, a respondent may identify a past colleague or a former classmate as a close contact. However, it is plausible that this self-reported connection is not captured in the population-scale social network data utilized, which only records formal ties active as of the date of the snapshot. It is crucial to take into account the varying time scales of these data sources when interpreting the results.

\subsection{Immigration Perception Index}
\label{sec:ipi}

Perceptions of immigration, the dependent variable in our study, are sourced from the survey data. In line with the established methodology of item response theory often used within the domain of psychometrics \parencite{Reckase1997, Jackman2009}, we calculate a continuous latent trait that captures views on immigration. This comprehensive index is constructed by combining a set of observable categorical variables, which are survey questions focused on the issue of immigration, following the approach used by \cite{Lancaster_2022}. As opposed to an additive index or an average of responses, the use of item response theory, more specifically, a graded response model \parencite{Samejima1997}, offers a more reliable weighting of questions in consideration. To construct a combined index, the graded response model estimates both item- and respondent-level parameters. An item-level estimate is a ‘discrimination parameter’ that indicates the ‘differential capability’ of a response item, i.e. it quantifies the extent to which an item conveys information about the underlying latent trait. Respondent-level ‘ability parameter’ places a person on the scale of a latent trait we are after, i.e. it captures one’s attitudes on immigration, and this estimated parameter is used in the study as an individual-level dependent variable.

Following ~\cite{Lancaster_2022}, we utilize all migration-related questions from LISS to calculate a comprehensive Immigration Perception Index. The Politics and Values study contains nine questions related to perceptions of immigration, the exhaustive list of which is presented in Supplementary Information~-~\nameref{para:xyz}. The selected questions focus on general acceptance of immigrants, acceptance of multiculturalism as well as immigrants’ rights and asylum policies. Questions are presented to the respondents as statements to which they indicate the level of agreement. Possible answers range on a scale from 1 to 5, where 1 is ‘fully disagree’, and 5 corresponds to ‘fully agree’. The index has a high estimated reliability of 0.79 as captured by Cronbach’s $\alpha$, similar to the results obtained by \cite{Lancaster_2022} who used the previous waves of the Politics and Values LISS study spanning from 2007 to 2019. This indicates that the constructed index has a relatively high representation of the latent trait of perceptions of immigration. The higher the value of the index, the more positive the views of the respondents towards immigration. More information on the distribution of the responses to the individual questions can be found in the Supplementary Information~\ref{tab:table_responses}. Some descriptive statistics of the created IPI index are presented in Supplementary Information~\ref{fig:fig6_distr}.

\subsection{Explanatory variables}
\label{sec:explanatory}

Below, we discuss the migrant background variable used to segment the respondents in our data, after which we cover social network measures and socio-demographic control variables used in our statistical analysis.

\textbf{Migration background variable.} The migration background\footnote{This study adheres to the definition of migrant background established prior to the revisions made by CBS in 2022, as detailed in \cite{CBS_ethn}. For an example of the usage of the new classification, see \cite{statistiek2024herkomstsegregatie}.} of an individual is derived from administrative registers. It is a nominal variable that takes the following possible values: Dutch, Moroccan, Turkish, Surinamese, Former Netherlands Antilles and Aruba, other western, and other non-western (see ethnic composition of the Dutch society in Supplementary Information~\ref{fig:fig1}-A). The migration background is defined based on the country of origin of an individual as well as one's parents. The migrant background of a person born outside the Netherlands who immigrated to the country, i.e. a first-generation migrant, is based on their country of birth. While such categories as Moroccan, Turkish, Surinamese, Former Netherlands Antilles, and Aruba strictly indicate the country/region of origin, the categories ‘other western’ and ‘other non-western’ group many countries together. ‘Other non-western’ includes all countries in South America, Africa, and Asia (except Indonesia and Japan). ‘Other western’ stands for all European countries (except Turkey), North America and Oceania, Indonesia, and Japan. Indonesia and Japan are included in the ‘other western’ category as migrants from those countries were to a large extent “born in the former Dutch East Indies as well as expatriates who were employed by Japanese companies along with their families” \parencite{CBS_nonwest}. Following CBS definitions, people who were born in the Netherlands and have at least one parent born abroad are considered second-generation migrants. In case it is only one parent who was born abroad, their children have the same migration background as a non-native Dutch parent. In case both parents were born abroad, their child’s migration background is defined based on the mother’s country of origin \parencite{CBS_ethn}.

While the existing categorization of ‘persons with a migration background’ is a helpful tool in accounting for intergenerational transmission of inequalities, it is necessary to acknowledge some of the exclusionary implications of the way it is designed \parencite{Will2019, Cranston2023, Kindler2014SocialNS}. The classification of migration background is centered around the country of birth of a person and their parents, and not the lived migration experience and self-identification of individuals \parencite{Will2019}. Furthermore, the aggregate categories such as ‘other western’ and ‘other non-western’ are broad and heterogeneous groups of populations. The selected aggregation strategy has been criticized as an “insufficient approach to gathering equal opportunities statistics”, in which unprivileged subgroups of migrants might be discarded in the context of these wider categories \parencite{Will2019, Yanow2015}.

\textbf{Social network measures.} Independent variables we constructed aim to capture close contact as well as the composition of the population-scale social opportunity structure of the respondents.

\begin{itemize}
    \item \textbf{Close contact:} in the LISS survey, respondents list direct close ties and perceived migration background of a respective alter. Based on this answer, we create the binary variable Close contact having the value 1 if at least one such contact with a migrant is reported, and 0 otherwise. 16.8\% of people report such a contact out of which 49.7\% are natives.
    
    \item We measure \textbf{Exposure} to migrants through the share of people with a migrant background (both first and second generation) within alters in the multilayer population-scale network. For 4,558 respondents we observe 381,693 alters in their social opportunity structures out of which 70,384 are migrants. We also measure Exposure within the network layers: family, household, school, work, and neighbor — which together as contexts compose the social opportunity structure of individuals.
    
    \item We also create a binary variable called \textbf{Multiplexity} to indicate if any immigrant contacts are present in one of the more social contexts for a respondent. We exclude family and household overlaps from both the binary variable and the counts because of their ubiquity. Out of 381,693 alters that are linked to survey respondents through register data, there are 581 double ties and 2 triple ties. Multiplex ties are present in social opportunity structures of 381 out of 4558 respondents, and only 67 respondents have multiplex ties with migrants.
    \item We count the number of above-defined \textbf{multiplex} ties.
    \item We use \textbf{degree}, that is, the total number of alters as a control variable both in the entire social opportunity structure and in each layer. 
\end{itemize}

\textbf{Socio-demographic variables.} Confounding factors are also sourced from the CBS’ register data. The covariates we account for include:
\begin{itemize}
    
\item \textbf{Age} corresponds to the age of respondents at the time of CBS’ data collection (October 2018). A prevalent hypothesis posits that individuals in older age groups are more likely to hold stronger anti-immigration attitudes. Nevertheless, a recent meta-analysis of drivers of attitudes toward immigration shows that the significance of continuous age is not consistent \parencite{Drazanova2022}. Instead, it is the broader generational, so-called “cohort” effect that proves to be more meaningful in predicting sentiment toward immigration \parencite{Schotte2018, Jeannet2019}. This supports including age as a control variable.

\item \textbf{Income.} While socio-economic status (such as income or occupational status) lacks significance in the majority of existing literature, we nevertheless include income \parencite{Drazanova2022}, operationalized as an equivalised combined household income, i.e. aggregate household income that is discounted by the size and composition of a household. To account for a skewed distribution of the income variable, we convert it into a percentile in the overall household income distribution of a population.

\item \textbf{Education} is a consistent and significant predictor of attitudes toward immigration, particularly within the subsample of North America and Western Europe, geographies typically studied in this context \parencite{Drazanova2022, Borgonovi2019, Hainmueller_Hiscox_2007}. The education covariate is an ordinal variable that captures the highest achieved level of education of an individual and takes the following values: primary, VMBO (secondary applied), HAVO/WVO (secondary scientific), MBO (vocational), WO (higher education). 

\item \textbf{Employment status} and \textbf{student status} are binary variables that indicate if a person is currently employed or in education, respectively. Clearly, this has a direct effect on the composition of the opportunity structure. 

\end{itemize}

\subsection{Statistical analysis} 
\label{sec:stats}
Statistical analysis of the association between social network composition and perceptions of immigration was done through ordinary least squares regression. An important consideration in the context of regression analysis is the sampling design of the survey. The way the survey in question is administered is by sampling households first and then surveying individuals who are the members of sampled households. Such a sampling design grants the necessity to cluster standard errors of regressions run on this sample at the level at which the sample was selected, i.e. at the household level \parencite{Abadie2023}. To illustrate the extent to which the outcome of interest, i.e. immigration attitudes, are related within different clusters in the population of interest (households, municipalities, etc.) we present the correlation of perceptions of immigration for respondents who belong to any of these clusters in Figure~\ref{fig:fig_correlation_ipi}-B. 

\textbf{Experimental setup.} 
We first develop a series of regressions with different variations of controls and their interactions to define the optimal set of confounding factors (Table \ref{table:controls}) to estimate the Immigration Perception Index. As a next step, we use the explanatory variables of interest - Close contact and Exposure - added to the controls in different combinations and settings.

Our first set of models aims to address the research question (\ref{word:rq1}) by comparing the importance of Close contact and Exposure in explaining perceptions of immigration for the subsamples of natives and migrants. Coefficients for second-generation migrants are estimated by incorporating a binary variable identifying this subgroup of the population, thereby reflecting a difference from the coefficient calculated for first-generation migrants:
\begin{itemize}
    
\item \textbf{Model~1} investigates the contribution of the Close contact.
\item \textbf{Model~2} explores the Exposure variable as the main explanatory variable of interest instead. 
\item \textbf{Model~3} adds a quadratic Exposure term to the Model~2.
\item \textbf{Model~4} combines all variables: Close Control, Exposure, and quadratic Exposure for the natives, and Close Control and Exposure for the migrant subgroups. 

\end{itemize}

We then test the contribution of Exposure for \textbf{Model~2} and \textbf{Model~3} in the different network layers, or social contexts, in order to shed light on the research question (\ref{word:rq2}). In the models testing the contribution of multiplex edges (in relation to the research question (\ref{word:rq3})), we use either Multiplexity or the number of multiplex ties with the previous control variables for the estimation.

\textbf{Reverse Causality. }
The selected research design, which prioritizes social opportunity structure, to a much smaller extent suffers from reverse causality compared to the more common setup that focuses on core networks. Unlike core networks, which consist of a narrow set of self-reported and self-selected ties, the social opportunity structure offers a broader, yet still meaningful, social exposure that is not entirely determined by an individual's choices. This broader spectrum of social exposure inherently includes various unplanned and incidental interactions and exposure that occur as a part of daily life, such as those with coworkers, or neighbors. Because these interactions are not solely based on personal choice or preference, they provide a less biased sample of social contacts. This randomness helps to mitigate the issue of reverse causality, where the outcome of interest (social perceptions of immigration) might influence the explanatory factor (composition of social opportunity structures). In this broader context, the directionality from social exposure to perceptions becomes clearer, as the individual's control over their exposure is significantly reduced.

\section*{Acknowledgements}

We are thankful for the collaboration with Statistics Netherlands colleagues Jan van der Laan, Edwin de Jonge, Gert Buiten, and Frank Pijpers. We would also like to thank the POPNET team~(\url{www.popnet.io}), especially Nicolas Soler, for helpful suggestions and discussions. We extend our gratitude to the PETGOV program group of the University of Amsterdam, particularly Annette Freyberg-Inan, Abbey Steele, Ruth Carlitz, Daniel Sloman, and Katharina Weber for their valuable suggestions.

The project has been funded by Platform Digitale Infrastructuur Social Sciences and Humanities~(\url{www.pdi-ssh.nl}). The LISS panel data were collected by Centerdata (Tilburg University, The Netherlands) through its MESS project funded by the Netherlands Organization for Scientific Research.

\printbibliography

\break

\section {Supplementary Information}
\beginsupplement

\subsection{Survey questions}
\label{para:xyz}

\paragraph{\textbf{cv19k104}} In the Netherlands, some people believe that immigrants are entitled to live here while retaining their own culture. Others feel that they should adapt entirely to Dutch culture. Where would you place yourself on a scale of 1 to 5, where 1 means that immigrants can retain their own culture and 5 means that they should adapt entirely? 
1 - immigrants can retain their own culture
5 - immigrants should adapt entirely to Dutch culture 
99 - I don't know


\begin{itemize}
    \item 1 - immigrants can retain their own culture
    \item 5 - immigrants should adapt entirely to Dutch culture 
    \item 99 - I don't know
\end{itemize}

What is your opinion on the following statements?

\textbf{cv19k116} It is good if society consists of people from different cultures. 

\textbf{cv19k117} It is difficult for a foreigner to be accepted in the Netherlands while retaining his/her own culture. 

\textbf{cv19k118} It should be made easier to obtain asylum in the Netherlands. 

\textbf{cv19k119} Legally residing foreigners should be entitled to the same social security as Dutch citizens. 

\textbf{cv19k120} There are too many people of foreign origin or descent in the Netherlands. 

\textbf{cv19k121} People of foreign origin or descent are not accepted in the Netherlands. 

\textbf{cv19k122} Some sectors of the economy can only continue to function because people of foreign origin or descent work there. 

\textbf{cv19k123} It does not help a neighborhood if many people of foreign origin or descent move in.

\begin{itemize}
    \item 1 - fully disagree 
    \item 2 - disagree 
    \item 3 - neither agree nor disagree 
    \item 4 - agree 
    \item 5 - fully agree
    \item 99 - I don't know

\end{itemize}

\subsection{Population and exposure to migrants}

\begin{figure}[ht!]
     \centering
     \includegraphics[width=\textwidth]{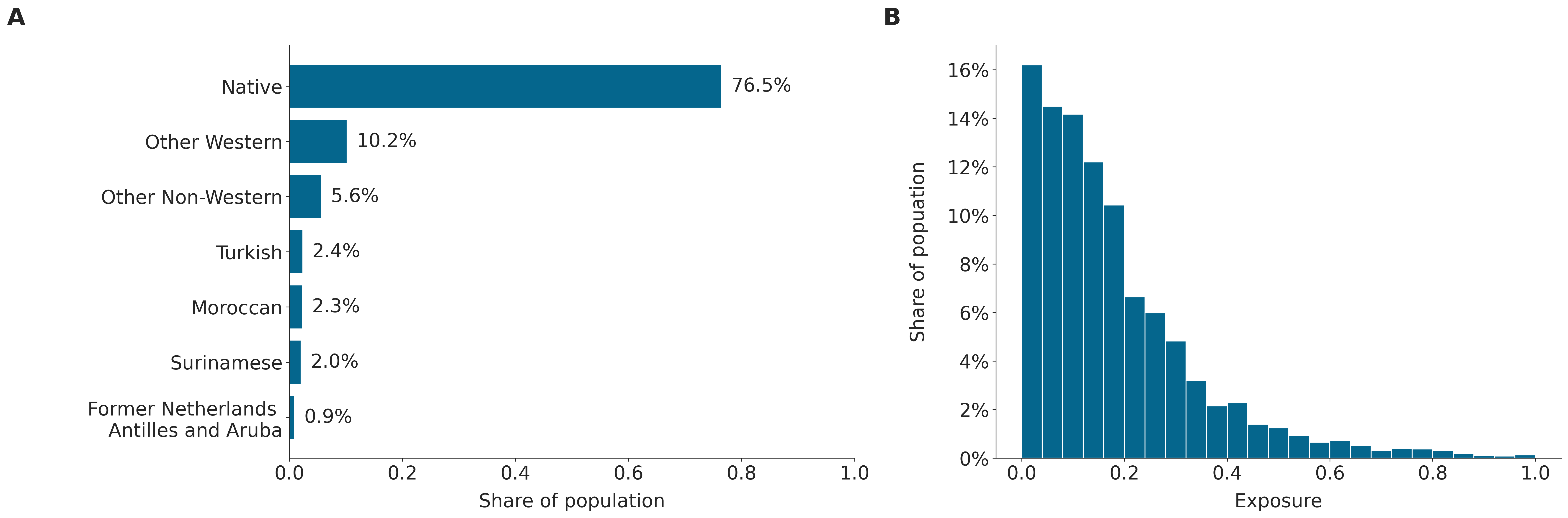}
     \caption{\textbf{Composition of the population with respect to migrant background:} (A) prevalence of each migrant group in the whole population, (B) distribution of the exposure to migrants in the social opportunity structures of survey respondents.}
     \label{fig:fig1}
 \end{figure}

Of the entire population, 76\% of the population is categorized as native Dutch in the administrative registers, and the remaining 24\% are migrants of both first and second generation with the most numerous groups being Turkish, Moroccan, Surinamese as well as migrants from Former Netherlands Antilles and Aruba (Figure~\ref{fig:fig1}-A). For the subset of survey respondents, the individual-level Exposure variable shows a wide unimodal distribution with a long right tail. Values range from no exposure to above 80\% (Figure~\ref{fig:fig1}-B), with an average of 18\%. 

\subsection{Immigration Perception Index}

The distribution of responses to constituting survey items is heterogeneous and presented below in Table~\ref{tab:table_responses}. 

\begin{table}[htbp]
\tiny 
  \centering
  \begin{tabularx}{\textwidth}{p{1.6cm} *{9}{>{\centering\arraybackslash}X}}
    \toprule
    \footnotesize & \tiny Immigrants can retain their own culture (1) up to immigrants should adapt entirely (5) & \tiny It is good if society consists of people from different cultures & \tiny It is difficult for a foreigner to be accepted in the Netherlands while retaining his/her own culture & \tiny It should be made easier to obtain asylum in the Netherlands & \tiny Legally residing foreigners should be entitled to the same social security as Dutch citizens & \tiny There are too many people of foreign origin or descent in the Netherlands & \tiny People of foreign origin or descent are not accepted in the Netherlands & \tiny Some sectors of the economy can only continue to function because people of foreign origin or descent work there & \tiny It does not help a neighborhood if many people of foreign origin or descent move in \\
    \midrule
    \tiny Fully disagree & 2\% & 3\% & 3\% & 22\% & 4\% & 7\% & 7\%& 6\% & 2\% \\
    \tiny Disagree & 7\% & 8\% & 17\% & 36\% & 10\% & 19\% & 38\% & 16\%& 11\% \\
     \tiny Neutral & 32\% & 25\% & 27\% & 26\% & 22\% & 34\% & 37\% &27\% &29\% \\
    \tiny Agree & 34\% & 45\% & 39\% & 7\% & 45\% & 23\% & 10\% & 38\% &37\% \\
    \tiny Fully agree & 15\% & 12\% & 7\% & 1\% & 11\% & 10\%  & 1\%& 5\%& 13\% \\
    \bottomrule
  \end{tabularx}
  \caption{\textbf{Views on immigration in Dutch society as captured by the LISS survey - Politics and Values, wave 11, dated April 2019.} }
  \label{tab:table_responses}%
\end{table}%

\begin{figure}[ht!]
    \centering
    \includegraphics[width=0.6\textwidth]{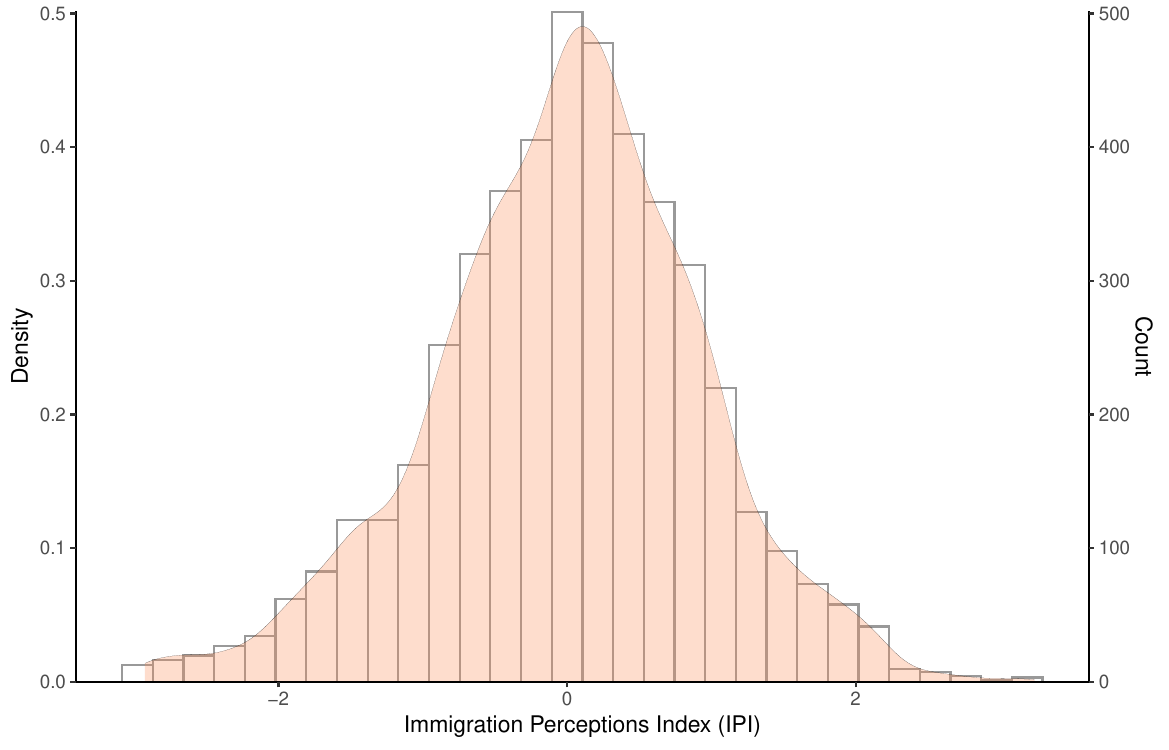}
    \caption{\textbf{Distribution of Immigration Perception Index (IPI)}}
    \label{fig:fig6_distr}
\end{figure}

The combined Immigration Perception Index based on all the nine survey items has close to normal distribution, however, it is not symmetric with values ranging between -2.92 and 3.24 (Figure~\ref{fig:fig6_distr}). Deviation from the symmetry of the distribution of the combined index is caused by the asymmetry and heavy tails of distributions of constituting response items. The overall distribution is slightly skewed in the direction of positive perceptions of immigration. Nevertheless, a large share of the respondents (44\%) fall into the range of -0.5 to 0.5 which indicates a relatively neutral sentiment towards immigration.

\begin{figure}[ht!]
    \centering
    \includegraphics[width=\textwidth]{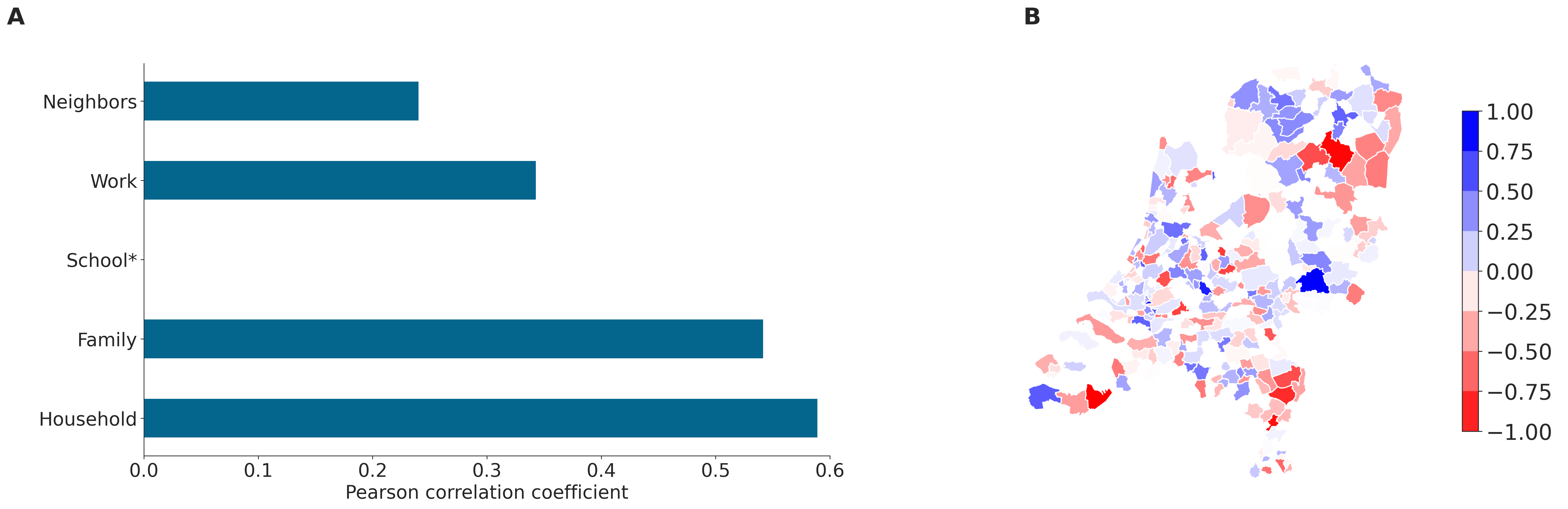}
    \caption{(A) \textbf{Pearson correlation of Immigration Perception Index} for respondents per network layer, (B) \textbf{Municipality-level average of Immigration Perception Index (source: LISS data)}}
    \label{fig:fig_correlation_ipi}
\end{figure}

Inspection of the interdependence of immigration attitudes between different respondents is implemented from the perspective of both the population-scale social network as well as geospatially. Figure~\ref{fig:fig_correlation_ipi}-A presents the correlation of the index for respondents that are connected in at least one of the layers of the population-scale social network. As the graph reveals, as expected, the correlation of the immigration attitudes between the household members that to a large extent overlap with family membership is strong at 0.59 and 0.54, respectively. The correlation of perceptions of immigration between work colleagues is moderate, at 0.34. Within the context of next-door neighbors, it is weaker, at 0.24. 


Aggregating the Immigration Perception Index on the municipality level (Figure~\ref{fig:fig_correlation_ipi}-B) reveals that all across the country, on average, the perceptions of immigration are neutral as the estimated municipality averages fall into a range between -1 and 1 while the index ranges from ~-3 to ~3. There are a few exceptions of municipalities that exhibit weakly positive or negative sentiments towards immigration. Overall, the spatial distribution of immigration attitudes is diverse, without a clear geographical clustering, yet within narrow boundaries of neutral or close to neutral immigration attitudes. 

\subsection{Regression tables}

In what follows we present the full regression summary tables.
\begin{itemize}
    \item Table~\ref{table:controls} presents a variety of regression models with IPI as the dependent variable and control variables only: age, income, education, migration background, employment status, and if a person is in education currently. 
    \item Tables \ref{tab:models_sos} and \ref{tab:models_sos_migrants} incorporate Close contact and Exposure variables for the subsamples of natives and migrants, respectively. These regression results are the basis of the Figures \ref{fig:fig3_reg_results} and \ref{fig:fig5_significant}-A.
    \item Tables \ref{table:coefficients} and \ref{table:coefficients_migrants} segment exposure by distinct social contexts for the subsamples of natives and migrants, respectively. These regression results are the basis of the Figures \ref{fig:fig4_layerwise} and \ref{fig:fig5_significant}-B.

\end{itemize}

\begin{table}
\footnotesize
\begin{tabular}{@{\extracolsep{5pt}}lD{.}{.}{-2} D{.}{.}{-2} D{.}{.}{-2} D{.}{.}{-2} D{.}{.}{-2} D{.}{.}{-2}} 
\\[-1.8ex]\hline 
\hline \\[-1.8ex] 
 & \multicolumn{6}{c}{\textit{Dependent variable:}} \\ 
\cline{2-7} 
\\[-1.8ex] & \multicolumn{6}{c}{Immigration Perception Index} \\ 
\\[-1.8ex] & \multicolumn{1}{c}{(1)} & \multicolumn{1}{c}{(2)} & \multicolumn{1}{c}{(3)} & \multicolumn{1}{c}{(4)} & \multicolumn{1}{c}{(5)} & \multicolumn{1}{c}{(6)}\\ 
\hline \\[-1.8ex] 
 Constant & -0.21^{*} & -0.46^{***} & -0.48^{***} & -0.34^{**} & -0.20 & -0.55^{***} \\ 
  & (0.10) & (0.10) & (0.11) & (0.12) & (0.12) & (0.12) \\ 

 Age & -0.001 & 0.003^{**} & 0.003^{**} & 0.001 & 0.002 & 0.002 \\ 
  & (0.001) & (0.001) & (0.001) & (0.001) & (0.001) & (0.001) \\ 

 Age * Employment \\
 status dummy &  &  &  & 0.003 & 0.002 & 0.002 \\ 
  &  &  &  & (0.002) & (0.002) & (0.002) \\ 

 Employment \\status dummy &  &  &  & -0.21^{*} & -0.16 & -0.16 \\ 
  &  &  &  & (0.10) & (0.10) & (0.09) \\ 

 Income & 0.0004 & 0.001 & 0.001 & 0.001 & 0.001^{*} & 0.001^{*} \\ 
  & (0.001) & (0.001) & (0.001) & (0.001) & (0.001) & (0.001) \\ 

 Education: Bachelor \\(baseline: primary) & 0.46^{***} & 0.50^{***} & 0.53^{***} & 0.55^{***} & 0.59^{***} & 0.61^{***} \\ 
  & (0.09) & (0.09) & (0.09) & (0.09) & (0.09) & (0.09) \\ 

 Education: Master/PhD \\(baseline: primary) & 0.65^{***} & 0.72^{***} & 0.74^{***} & 0.76^{***} & 0.78^{***} & 0.80^{***} \\ 
  & (0.09) & (0.09) & (0.10) & (0.10) & (0.10) & (0.10) \\ 

 Missing education dummy & 0.11 & 0.11 & 0.12 & 0.14 & 0.18^{*} & 0.20^{*} \\ 
  & (0.08) & (0.08) & (0.09) & (0.09) & (0.09) & (0.09) \\ 

 Education: Secondary \\(baseline: primary) & 0.14 & 0.08 & 0.11 & 0.11 & 0.13 & 0.15 \\ 
  & (0.09) & (0.09) & (0.10) & (0.10) & (0.10) & (0.10) \\ 

 Education: Vocational \\(baseline: primary) & 0.16 & 0.17 & 0.17 & 0.19^{*} & 0.23^{*} & 0.24^{**} \\ 
  & (0.09) & (0.09) & (0.09) & (0.09) & (0.09) & (0.09) \\ 

 Education: Bachelor (baseline: \\primary) * Currently in education &  &  & -0.22 & -0.23 & -0.18 & -0.23 \\ 
  &  &  & (0.28) & (0.28) & (0.28) & (0.28) \\ 

 Education: Master/PhD (baseline: \\primary) * Currently in education &  &  & -0.58 & -0.56 & -0.49 & -0.53 \\ 
  &  &  & (0.45) & (0.45) & (0.45) & (0.45) \\ 

 Missing education * Currently \\in education &  &  &  &  &  &  \\ 
  &  &  &  &  &  &  \\ 

 Education: Secondary (baseline:\\ primary) * Currently in education &  &  & -0.18 & -0.16 & -0.13 & -0.16 \\ 
  &  &  & (0.28) & (0.28) & (0.28) & (0.27) \\ 

 Education: Vocational (baseline:\\ primary) * Currently in education &  &  & -0.08 & -0.07 & -0.02 & -0.06 \\ 
  &  &  & (0.27) & (0.27) & (0.26) & (0.26) \\ 

 Currently in education &  & 0.44^{***} & 0.58^{*} & 0.56^{*} & 0.52^{*} & 0.58^{*} \\ 
  &  & (0.06) & (0.26) & (0.26) & (0.26) & (0.26) \\ 

 Native Dutch dummy &  &  &  &  & -0.33^{***} &  \\ 
  &  &  &  &  & (0.04) &  \\ 

 1$^{st}$ gen (baseline: native) &  &  &  &  &  & 0.44^{***} \\ 
  &  &  &  &  &  & (0.05) \\ 

 2$^{nd}$ gen (baseline: native) &  &  &  &  &  & 0.23^{***} \\ 
  &  &  &  &  &  & (0.05) \\ 

\hline \\[-1.8ex] 
Observations & \multicolumn{1}{c}{4,538} & \multicolumn{1}{c}{4,538} & \multicolumn{1}{c}{4,538} & \multicolumn{1}{c}{4,538} & \multicolumn{1}{c}{4,538} & \multicolumn{1}{c}{4,538} \\ 
R$^{2}$ & \multicolumn{1}{c}{0.04} & \multicolumn{1}{c}{0.06} & \multicolumn{1}{c}{0.06} & \multicolumn{1}{c}{0.06} & \multicolumn{1}{c}{0.07} & \multicolumn{1}{c}{0.08} \\ 
Adjusted R$^{2}$ & \multicolumn{1}{c}{0.04} & \multicolumn{1}{c}{0.05} & \multicolumn{1}{c}{0.05} & \multicolumn{1}{c}{0.05} & \multicolumn{1}{c}{0.07} & \multicolumn{1}{c}{0.07} \\ 
\hline 
\hline \\[-1.8ex] 
\textit{Note:}  & \multicolumn{6}{r}{$^{*}$p$<$0.05; $^{**}$p$<$0.01; $^{***}$p$<$0.001} \\ 
\end{tabular} 
\caption{\textbf{Baseline models with controls only.}} 
\label{table:controls} 
\end{table} 

\begin{table}[!htbp] \centering 
 \footnotesize
\begin{tabular}{@{\extracolsep{5pt}}lD{.}{.}{-2} D{.}{.}{-2} D{.}{.}{-2} D{.}{.}{-2} } 
\\[-1.8ex]\hline 
\hline \\[-1.8ex] 
 & \multicolumn{3}{c}{\textit{Dependent variable:}} \\ 
\cline{2-4} 
\\[-1.8ex] & \multicolumn{3}{c}{Immigration Perception Index} \\ 
\\[-1.8ex] & \multicolumn{1}{c}{Model~1 - Natives} & \multicolumn{1}{c}{Model~2 - Natives} &  \multicolumn{1}{c}{Model~3 - Natives} & \multicolumn{1}{c}{Model~4 - Natives}\\ 
\hline \\[-1.8ex] 
 Close contact & \cellcolor{bleucite!30}0.34^{***} &  & & \cellcolor{bleucite!30}0.32^{***} \\ 
  & (0.06) &  & &  (0.06) \\ 

 Exposure & & 0.27 & \cellcolor{bleucite!30}0.99^{**} & 0.27 \\ 
  & & (0.14)& (0.35) & (0.38) \\ 

 Exposure$^{2}$ & & & \cellcolor{bleucite!30}-1.51^{*} & -0.13 \\ 
  &  & & (0.57) & (0.73) \\ 

 Degree & 0.0000 &-0.0002 & -0.0003 & -0.0000 \\ 
  & (0.0004) & (0.0003)& (0.0003) & (0.0004) \\ 

\hline \\[-1.8ex] 
Observations & \multicolumn{1}{c}{2,884} &\multicolumn{1}{c}{3,750} & \multicolumn{1}{c}{3,750} & \multicolumn{1}{c}{2,884} \\ 
R$^{2}$ & \multicolumn{1}{c}{0.07} & \multicolumn{1}{c}{0.07}&\multicolumn{1}{c}{0.07} & \multicolumn{1}{c}{0.07} \\ 
Adjusted R$^{2}$ & \multicolumn{1}{c}{0.06} &\multicolumn{1}{c}{0.06}  &\multicolumn{1}{c}{0.06} & \multicolumn{1}{c}{0.06} \\ 
\hline 
\hline \\[-1.8ex] 
\textit{Note:} & & \multicolumn{3}{r}{$^{*}$p$<$0.1; $^{**}$p$<$0.05; $^{***}$p$<$0.01} \\ 
\end{tabular} 
 \caption{\textbf{Multivariate regression results estimating the relationship between the Immigration Perception Index and Close contact as well as Exposure} to migrants for the subsample of native respondents.  Standard errors are clustered at the level of households and presented in parentheses.} 
  \label{tab:models_sos} 
\end{table} 

\begin{table}[!htbp] \centering 
\footnotesize
\begin{tabular}{@{\extracolsep{5pt}}lD{.}{.}{-2} D{.}{.}{-2} D{.}{.}{-2} D{.}{.}{-2} } 
\\[-1.8ex]\hline 
\hline \\[-1.8ex] 
 & \multicolumn{4}{c}{\textit{Dependent variable:}} \\ 
\cline{2-5} 
\\[-1.8ex] & \multicolumn{4}{c}{Immigration Perception Index} \\ 
\\[-1.8ex] & \multicolumn{1}{c}{Model~1 - Migrants} & \multicolumn{1}{c}{Model~2 - Migrants} & \multicolumn{1}{c}{Model~3 - Migrants} & \multicolumn{1}{c}{Model~4 - Migrants}\\ 
\hline \\[-1.8ex] 
 Close contact & \cellcolor{bleucite!30}0.24^{**} &  &  & \cellcolor{bleucite!30} 0.23^{**}  \\ 
  & (0.09) &  &  & (0.09)  \\ 

 Exposure &  & \cellcolor{bleucite!30}0.48^{***} & 0.46 & 0.17 \\ 
  &  & (0.18) & (0.62) & (0.20) \\ 

 Exposure$^{2}$ &  &  & 0.02 & \\ 
  &  &  & (0.69)&  \\ 
 
 2$^{nd}$ generation & -0.13 & \cellcolor{bleucite!30}-0.19^{***} & \cellcolor{bleucite!30}-0.19^{***} & \\ 
  &(0.08)& (0.07) & (0.07)& -0.12 \\ 
  & & & &(0.08) \\ 
 Degree & 0.001 & 0.001^{*} & 0.001 & 0.001 \\ 
  & (0.001) & (0.001) & (0.001) & (0.001) \\ 

\hline \\[-1.8ex] 
Observations & \multicolumn{1}{c}{537} & \multicolumn{1}{c}{788} & \multicolumn{1}{c}{788} & \multicolumn{1}{c}{537} \\ 
R$^{2}$ & \multicolumn{1}{c}{0.09} & \multicolumn{1}{c}{0.08} & \multicolumn{1}{c}{ } & \multicolumn{1}{c}{0.09} \\ 
Adjusted R$^{2}$ & \multicolumn{1}{c}{0.06} & \multicolumn{1}{c}{0.06} & \multicolumn{1}{c}{ } & \multicolumn{1}{c}{0.06} \\ 
\hline 
\hline \\[-1.8ex] 
\textit{Note:}  & \multicolumn{4}{r}{$^{*}$p$<$0.1; $^{**}$p$<$0.05; $^{***}$p$<$0.01} \\ 
\end{tabular} 

  \caption{\textbf{Multivariate regression results estimating the relationship between the Immigration Perception Index and Close contact as well as Exposure} to migrants for the subsample of migrant respondents.  Standard errors are clustered at the level of households and presented in parentheses.} 
  \label{tab:models_sos_migrants} 
\end{table} 

\begin{landscape}

\begin{table}
\footnotesize
\begin{center}
\begin{tabular}{l c c c c c c c c c c c c}
\hline
 & Model~1 & Model~2 & Model~3 & Model~4 & Model~5 & Model~6 & Model~7 & Model~8 & Model~9 & Model~10 & Model~11 & Model~12 \\
\hline
Exposure H  &  \cellcolor{bleucite!30}$0.12^{**}$ &             &          &          &          & $0.14^{***}$ & $0.75^{*}$  &            &          &          &             & $0.44$     \\
                       & $(0.04)$    &             &          &          &          & $(0.03)$     & $(0.38)$    &            &          &          &             & $(0.38)$   \\
Degree H   & $-0.03^{*}$ &             &          &          &          &              & $-0.04^{*}$ &            &          &          &             &            \\
                       & $(0.02)$    &             &          &          &          &              & $(0.02)$    &            &          &          &             &            \\
Exposure F  &             &  \cellcolor{bleucite!30}$0.23^{**}$ &          &          &          & $0.16$       &             & $0.45^{*}$ &          &          &             & $0.26$     \\
                       &             & $(0.08)$    &          &          &          & $(0.08)$     &             & $(0.23)$   &          &          &             & $(0.23)$   \\
Degree F   &             & $-0.00$     &          &          &          &              &             & $-0.00$    &          &          &             &            \\
                       &             & $(0.00)$    &          &          &          &              &             & $(0.00)$   &          &          &             &            \\
Exposure S  &             &             & $0.23$   &          &          & $0.19$       &             &            & $0.69$   &          &             & $0.44$     \\
                       &             &             & $(0.32)$ &          &          & $(0.31)$     &             &            & $(0.76)$ &          &             & $(0.76)$   \\
Degree S   &             &             & $-0.00$  &          &          &              &             &            & $-0.00$  &          &             &            \\
                       &             &             & $(0.00)$ &          &          &              &             &            & $(0.00)$ &          &             &            \\
Exposure W  &             &             &          & $0.08$   &          & $0.01$       &             &            &          & $0.25$   &             & $0.01$     \\
                       &             &             &          & $(0.11)$ &          & $(0.11)$     &             &            &          & $(0.28)$ &             & $(0.29)$   \\
Degree W   &             &             &          & $0.00$   &          &              &             &            &          & $0.00$   &             &            \\
                       &             &             &          & $(0.00)$ &          &              &             &            &          & $(0.00)$ &             &            \\
Exposure N  &             &             &          &          & $0.13$   & $0.08$       &             &            &          &          &  \cellcolor{bleucite!30}$0.52^{**}$ & $0.36^{*}$ \\
                       &             &             &          &          & $(0.07)$ & $(0.07)$     &             &            &          &          & $(0.18)$    & $(0.19)$   \\
Degree N   &             &             &          &          & $0.00$   &              &             &            &          &          & $-0.00$     &            \\
                       &             &             &          &          & $(0.00)$ &              &             &            &          &          & $(0.00)$    &            \\
Degree         &             &             &          &          &          & $-0.00$      &             &            &          &          &             & $-0.00$    \\
                       &             &             &          &          &          & $(0.00)$     &             &            &          &          &             & $(0.00)$   \\
Exposure H$^{2}$ &             &             &          &          &          &              & $-0.64$     &            &          &          &             & $-0.30$    \\
                       &             &             &          &          &          &              & $(0.38)$    &            &          &          &             & $(0.38)$   \\
Exposure F$^{2}$ &             &             &          &          &          &              &             & $-0.27$    &          &          &             & $-0.15$    \\
                       &             &             &          &          &          &              &             & $(0.26)$   &          &          &             & $(0.27)$   \\
Exposure S$^{2}$ &             &             &          &          &          &              &             &            & $-0.56$  &          &             & $-0.34$    \\
                       &             &             &          &          &          &              &             &            & $(0.86)$ &          &             & $(0.85)$   \\
Exposure W$^{2}$ &             &             &          &          &          &              &             &            &          & $-0.22$  &             & $-0.03$    \\
                       &             &             &          &          &          &              &             &            &          & $(0.33)$ &             & $(0.33)$   \\
Exposure N$^{2}$ &             &             &          &          &          &              &             &            &          &          &  \cellcolor{bleucite!30}$-0.51^{*}$ & $-0.36$    \\
                       &             &             &          &          &          &              &             &            &          &          & $(0.23)$    & $(0.22)$   \\
\hline \\[-1.8ex] 
Observations & \multicolumn{1}{c}{3,750} & \multicolumn{1}{c}{3,750} & \multicolumn{1}{c}{3,750} & \multicolumn{1}{c}{3,750}& \multicolumn{1}{c}{3,750}& \multicolumn{1}{c}{3,750}& \multicolumn{1}{c}{3,750}& \multicolumn{1}{c}{3,750}& \multicolumn{1}{c}{3,750}& \multicolumn{1}{c}{3,750}& \multicolumn{1}{c}{3,750}& \multicolumn{1}{c}{3,750} \\ 

R$^{2}$ & \multicolumn{1}{c}{0.07} & \multicolumn{1}{c}{0.06} & \multicolumn{1}{c}{0.06} & \multicolumn{1}{c}{0.06} & \multicolumn{1}{c}{0.06} & \multicolumn{1}{c}{0.07}& \multicolumn{1}{c}{0.07}& \multicolumn{1}{c}{0.07}& \multicolumn{1}{c}{0.06}& \multicolumn{1}{c}{0.06}& \multicolumn{1}{c}{0.07}& \multicolumn{1}{c}{0.07} \\ 
Adjusted R$^{2}$ & \multicolumn{1}{c}{0.07} & \multicolumn{1}{c}{0.06} & \multicolumn{1}{c}{0.06} & \multicolumn{1}{c}{0.06} & \multicolumn{1}{c}{0.06}  & \multicolumn{1}{c}{0.07} & \multicolumn{1}{c}{0.07}& \multicolumn{1}{c}{0.06}& \multicolumn{1}{c}{0.06}& \multicolumn{1}{c}{0.06}& \multicolumn{1}{c}{0.06}& \multicolumn{1}{c}{0.07}\\ 
\hline 

\tiny
\textit {Note:} & & &  & &  & &  & &  \multicolumn{4}{r}{ \tiny $^{*}$p$<$0.1; $^{**}$p$<$0.05; $^{***}$p$<$0.01} \\ 
\end{tabular}
\caption{\textbf{Multivariate regression results estimating the relationship between IPI and Exposure to migrants in different contexts} of population-scale social network for the subsample of natives. Standard errors are heteroskedasticity-consistent.}
\label{table:coefficients}
\end{center}
\end{table}

\begin{table}
\footnotesize
\begin{center}
\begin{tabular}{l c c c c c c c c c c c c}
\hline
 & Model~1 & Model~2 & Model~3 & Model~4 & Model~5 & Model~6 & Model~7 & Model~8 & Model~9 & Model~10 & Model~11 & Model~12 \\
\hline
Exposure H  & \cellcolor{bleucite!30}$0.26^{***}$ &              &          &          &             & $0.10$       & \cellcolor{bleucite!30}$-1.09^{*}$ &              &          &          &            & $-0.80$    \\
                       & $(0.08)$     &              &          &          &             & $(0.09)$     & $(0.45)$    &              &          &          &            & $(0.43)$   \\
Degree H   & $0.05^{*}$   &              &          &          &             &              & $0.08^{**}$ &              &          &          &            &            \\
                       & $(0.02)$     &              &          &          &             &              & $(0.02)$    &              &          &          &            &            \\
Exposure F  &              & \cellcolor{bleucite!30}$0.55^{***}$ &          &          &             & $0.39^{***}$ &             & $0.34$       &          &          &            & $0.48$     \\
                       &              & $(0.10)$     &          &          &             & $(0.11)$     &             & $(0.43)$     &          &          &            & $(0.43)$   \\
Degree F   &              & $0.01^{***}$ &          &          &             &              &             & $0.01^{***}$ &          &          &            &            \\
                       &              & $(0.00)$     &          &          &             &              &             & $(0.00)$     &          &          &            &            \\
Exposure S  &              &              & $-0.37$  &          &             & $-0.64$      &             &              & $0.16$   &          &            & $-0.43$    \\
                       &              &              & $(0.37)$ &          &             & $(0.36)$     &             &              & $(1.41)$ &          &            & $(1.31)$   \\
Degree S   &              &              & $0.00$   &          &             &              &             &              & $0.00$   &          &            &            \\
                       &              &              & $(0.00)$ &          &             &              &             &              & $(0.00)$ &          &            &            \\
Exposure W  &              &              &          & $-0.08$  &             & $-0.25$      &             &              &          & $0.75$   &            & $0.09$     \\
                       &              &              &          & $(0.23)$ &             & $(0.25)$     &             &              &          & $(0.67)$ &            & $(0.67)$   \\
Degree W   &              &              &          & $0.00$   &             &              &             &              &          & $0.00$   &            &            \\
                       &              &              &          & $(0.00)$ &             &              &             &              &          & $(0.00)$ &            &            \\
Exposure N  &              &              &          &          & \cellcolor{bleucite!30}$0.35^{**}$ & $0.23$       &             &              &          &          & $0.10$     & $0.06$     \\
                       &              &              &          &          & $(0.13)$    & $(0.15)$     &             &              &          &          & $(0.41)$   & $(0.42)$   \\
Degree N   &              &              &          &          & $0.01$      &              &             &              &          &          & $0.01^{*}$ &            \\
                       &              &              &          &          & $(0.00)$    &              &             &              &          &          & $(0.00)$   &            \\
Degree         &              &              &          &          &             & $0.00$       &             &              &          &          &            & $0.00$     \\
                       &              &              &          &          &             & $(0.00)$     &             &              &          &          &            & $(0.00)$   \\
Exposure H$^{2}$ &              &              &          &          &             &              & \cellcolor{bleucite!30}$1.34^{**}$ &              &          &          &            & $0.89^{*}$ \\
                       &              &              &          &          &             &              & $(0.44)$    &              &          &          &            & $(0.42)$   \\
Exposure F$^{2}$ &              &              &          &          &             &              &             & $0.18$       &          &          &            & $-0.10$    \\
                       &              &              &          &          &             &              &             & $(0.35)$     &          &          &            & $(0.36)$   \\
Exposure S$^{2}$ &              &              &          &          &             &              &             &              & $-0.58$  &          &            & $-0.35$    \\
                       &              &              &          &          &             &              &             &              & $(1.44)$ &          &            & $(1.31)$   \\
Exposure W$^{2}$ &              &              &          &          &             &              &             &              &          & $-0.95$  &            & $-0.37$    \\
                       &              &              &          &          &             &              &             &              &          & $(0.81)$ &            & $(0.78)$   \\
Exposure N$^{2}$ &              &              &          &          &             &              &             &              &          &          & $0.29$     & $0.19$     \\
                       &              &              &          &          &             &              &             &              &          &          & $(0.45)$   & $(0.46)$   \\

\hline \\[-1.8ex] 
Observations & \multicolumn{1}{c}{788} & \multicolumn{1}{c}{788} & \multicolumn{1}{c}{788} & \multicolumn{1}{c}{788}& \multicolumn{1}{c}{788}& \multicolumn{1}{c}{788}& \multicolumn{1}{c}{788}& \multicolumn{1}{c}{788}& \multicolumn{1}{c}{788}& \multicolumn{1}{c}{788}& \multicolumn{1}{c}{788}& \multicolumn{1}{c}{788} \\ 

R$^{2}$ & \multicolumn{1}{c}{0.06} & \multicolumn{1}{c}{0.09} & \multicolumn{1}{c}{0.04} & \multicolumn{1}{c}{0.04} & \multicolumn{1}{c}{0.06} & \multicolumn{1}{c}{0.06}& \multicolumn{1}{c}{0.07}& \multicolumn{1}{c}{0.09}& \multicolumn{1}{c}{0.04}& \multicolumn{1}{c}{0.05}& \multicolumn{1}{c}{0.06}& \multicolumn{1}{c}{0.09} \\ 
Adjusted R$^{2}$ & \multicolumn{1}{c}{0.05} & \multicolumn{1}{c}{0.07} & \multicolumn{1}{c}{0.03} & \multicolumn{1}{c}{0.03} & \multicolumn{1}{c}{0.04}  & \multicolumn{1}{c}{0.07} & \multicolumn{1}{c}{0.06}& \multicolumn{1}{c}{0.07}& \multicolumn{1}{c}{0.03}& \multicolumn{1}{c}{0.03}& \multicolumn{1}{c}{0.04}& \multicolumn{1}{c}{0.07}\\ 

\hline
\tiny
\textit  {Note:} & & &  & &  & &  & &  \multicolumn{4}{r}{ \tiny $^{*}$p$<$0.1; $^{**}$p$<$0.05; $^{***}$p$<$0.01} \\ 
\end{tabular}

\caption{\textbf{Multivariate regression results estimating the relationship between IPI and Exposure to migrants in different contexts} of population-scale social network for the subsample of migrants. Standard errors are heteroskedasticity-consistent.}
\label{table:coefficients_migrants}
\end{center}
\end{table}

\end{landscape}


\end{document}